\documentclass[twocolumn,prl,twocolumn,english,superscriptaddress,floatfix]{revtex4-1}
\def \FigPath {./}

\usepackage{graphicx}
\usepackage{float}
\usepackage{color}
\usepackage{bm,amsmath,amssymb}
\usepackage{soul}
\usepackage[T1]{fontenc}

\usepackage[utf8]{inputenc}
\usepackage{hyperref}
\hypersetup{
  colorlinks = false,
  urlcolor = black,
}

\usepackage{braket}
\usepackage{xfrac}
\usepackage{verbatim}
\usepackage{lipsum}
\usepackage{cancel}
\usepackage{babel}
\usepackage{relsize}

\usepackage{amstext}
\usepackage{bbold}
\usepackage{esint}

\newcommand{\tr}[1]{\mathrm{tr}[#1]}
\newcommand{\eye}{\hat{1}}
\newcommand{\lnR}{\mathcal{Q}}

\newcommand{\redline}[1]{\textcolor{black}{#1}}

\begin{document}

\title{Characterizing a statistical arrow of time in quantum measurement dynamics}

\author{P. M. Harrington*$^\dagger$}
\affiliation{Department of Physics, Washington University, St.\ Louis, Missouri 63130}
\author{D. Tan*} 
\affiliation{Department of Physics, Washington University, St.\ Louis, Missouri 63130}
\affiliation{Shenzhen Institute for Quantum Science and Engineering and Department of Physics, Southern University of Science and Technology, Shenzhen 518055, People’s Republic of China}
\author{M. Naghiloo}
\affiliation{Department of Physics, Washington University, St.\ Louis, Missouri 63130}
\author{K. W. Murch$^\dagger$}
\affiliation{Department of Physics, Washington University, St.\ Louis, Missouri 63130}
\affiliation{Institute for Materials Science and Engineering, St.\ Louis, Missouri 63130}

\date{\today}

\begin{abstract}
In both thermodynamics and quantum mechanics the arrow of time is characterized by the statistical likelihood of physical processes. We characterize this arrow of time for the continuous quantum measurement dynamics of a superconducting qubit. By experimentally tracking individual weak measurement trajectories, we compare the path probabilities of forward and backward-in-time evolution to develop an arrow of time statistic associated with measurement dynamics. We compare the statistics of individual trajectories to ensemble properties showing that the measurement dynamics obeys both detailed and integral fluctuation theorems thus establishing the consistency between microscopic and macroscopic measurement dynamics.
\end{abstract}

\maketitle

The entanglement between a quantum system and its environment can be harnessed for indirect measurements, where measurements on the environment alone convey information and induce backaction on the system \cite{bragbook,wisebook}. Because the outcomes of measurements on quantum systems are inherently probabilistic, the outcomes of measurements on the environment have a random character and are statistically described by the quantum state as a model parameter. As the quantum state informs a predictive model of environment fluctuations, experimental measurements on the environment can serve as a predictor for the quantum state. In the fashion of Bayesian inference, quantum state tracking then consists of estimating model parameters conditioned on experimentally detected environment fluctuations resulting in a conditional stochastic evolution of the quantum state---a quantum trajectory \cite{wisebook,jacobsbook}.

Recent experimental capabilities in cavity quantum electrodynamics have enabled high efficiency sampling of environment fluctuations and tracking of individual quantum trajectories \cite{guer07,murc13,roch14,camp16,nagh16}, including statistical properties of these trajectories \cite{chan13,webe14,jord16,aree17,nagh17caustic}. 
These quantum trajectories bear a conceptual similarity to classical stochastic trajectories of particles that interact with a thermal reservoir. For such classical trajectories, entropy production can be characterized by tracking the evolution of single particles and comparing the probability density for forward versus time reversed trajectories \cite{croo99,croo00, seif05, harr07,horo11}. \redline{Experiments in classical systems \cite{lip02,wan02,tiet06,espo07,bli06,spec07,utsu10,kung12,sair12,kosk13,hoan18} have verified that these entropy measures satisfy fundamental fluctuation theorems that relate microscopic dynamics to ensemble behavior \cite{jarz97,croo98,espo10,jar11,seif12,cil13,manz18}. More broadly, these are related to fluctuation theorems for work distributions which have been extended to quantum systems \cite{kurc01, muka03, camp09, deff11, mori11, chet12, horo13}. There have been several proposals for experimental tests \cite{quan06, quan07,dorn13,mazz13,camp13,gool14,ronc14,chia15} with recent experimental results in closed quantum systems \cite{bata15, an15}.  In contrast, open quantum systems present new phenomena associated measurement backaction \cite{alon16,wata14,yi14,camp11,wata14_generalized,elou17_role,beno18,elou17_extracting,elou18,buff18}. In this letter, we characterize the entropy production of an open quantum system with individual quantum measurement trajectories \cite{espo09,camp11,elo17,alon16,nagh17}, using information entropy measures to characterize a statistical arrow of time in quantum measurement.} We show how a statistical arrow of time is revealed by path probabilities of forward versus time reversed quantum trajectories \cite{Bookschulman, Dressel2017,kaon17,sree18}. As in the case of classical trajectories, these probability densities satisfy a fluctuation theorem that is consistent with the correspondence between microscopic dynamics and ensemble behavior.

Our experiment focuses on a paradigmatic system of quantum measurement consisting of a pseudo spin-half system coupled to a single mode of the electromagnetic field (Fig.~1a) \cite{murc13,webe14}. The two lowest levels of a transmon circuit \cite{koch07,paik113D} give a qubit transition frequency $\omega_q/2\pi=4.01$ GHz, and coupling to a microwave cavity results in a dispersive interaction given by the  interaction Hamiltonian $H_\mathrm{int}=-\chi a^\dagger\!a\sigma_z$, where $\chi/2\pi=-0.6$ MHz is the dispersive coupling rate, $a^\dagger\!a$ is the number operator for the cavity mode at frequency $\omega_c/2\pi=6.8316$ GHz, and $\sigma_z$ is the Pauli operator that commutes with the qubit Hamiltonian. Qubit measurement occurs when a microwave tone probes the cavity resonance, acquiring a qubit-state-dependent phase shift. The shift on the cavity resonance $2|\chi|$ is small compared to the cavity linewidth $\kappa/2\pi = 9.0\,\text{MHz}$, endowing the measurement tone with a relatively small qubit-state-dependent phase shift. By virtue of this qubit--cavity interaction, the qubit state is correlated to a single field quadrature of the microwave probe, which is subsequently amplified by a near-quantum-limited Josephson parametric amplifier \cite{cast08,hatr11para} operating in phase sensitive mode.

\begin{figure}
\begin{center}
\includegraphics[width = 0.48\textwidth]{\FigPath 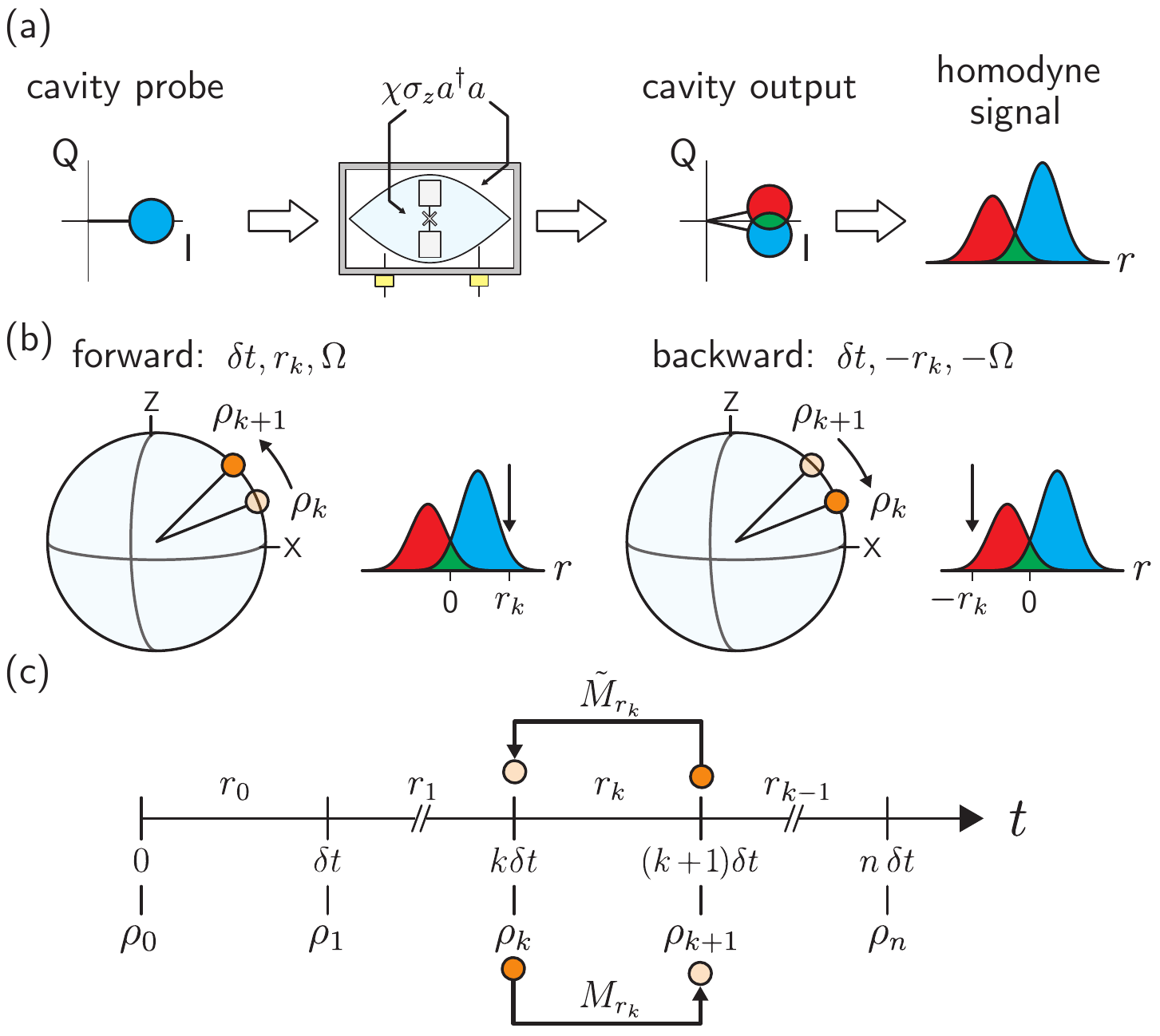}
\end{center}
\caption{\label{fig1}\small{Forward and reversed trajectories of a continuously monitored superconducting qubit. (a) The experiment setup consists of a superconducting qubit dispersively coupled to the fundamental mode of a waveguide cavity. The output signal reflected from the cavity acquires a qubit-state-dependent phase shift. (b) In a single update step, a measurement record $r_k$ of duration $\delta t$ from a continuous cavity probe induces backaction on the quantum state.  Upon time reversal of this update step, the state responds to backaction of a measurement result of opposite sign $-r_k$ by returning to the initial state. (c) Schematic of the state and measurement labels for forward ($M_{r_k}$) and backward  ($\tilde{M}_{r_k}$) state update procedures.}}
\end{figure}

The amplified quadrature is downconverted to DC and digitized into timesteps $\{t_k\}_{k=0}^{k=n}$ to obtain a set of measurement records $\{r_k\}_{k=0}^{k=n-1}$. From these measurement records, we reconstruct piecewise continuous trajectories, with each individual trajectory captured by a time-series of density operators $\{\rho_k\}_{k=0}^{k=n}$ \cite{breu07}. Informed by successive measurement records in a time-series, we use an iterative update scheme to infer the qubit state and realize a single trajectory. The dynamics of the trajectory results from the impression each stochastic measurement record $r_k$ has on our state-of-knowledge $\rho_k$.

The statistics of the measurement record and the dynamics imparted on the state are described by the positive \redline{operator-valued} measure (POVM) \cite{jaco06}, 
\begin{eqnarray}
M_{r_k} = \bigg(\frac{\delta t}{2 \pi\tau}\bigg)^{1/4}\exp\bigg[\!
-\frac{\delta t}{4\tau}(r_k\eye-\sigma_z)^2\bigg]
\end{eqnarray}
where the measurement strength is the product of the signal integration duration $\delta t=t_{k+1}-t_{k}=16\,\text{ns}$ and the measurement rate $1/\tau = 1.97\,\mu\text{s}^{-1}$. In addition to measurement, the evolution includes a dynamics due to a resonant Rabi drive by the Hamiltonian $H/\hbar=\Omega\sigma_y/2$ with $\Omega/2\pi = 2.16\,\text{MHz}$, which is in a rotating frame of the qubit transition. The POVM provides a state update conditioned on the measurement record from the relation $\rho_{k+1} = M_{r_k} \rho_k M_{r_k}^\dagger /\tr{M_{r_k} \rho_k M_{r_k}^\dagger}$, where the probability density of the measurement outcome is given by $P(r_k|\rho_k)\,dr_k = \tr{M_{r_k} \rho_k M_{r_k}^\dagger}\,dr_k$. By this probability density, and the associated information entropy, we statistically examine time reversal in the measurement process by comparing the likelihood of quantum trajectories that are ordered forward versus backward in time.

\redline{The notion that the quantum measurement process can be reversed stems from studies of `measurement undoing' \cite{koro06}, where weak measurements can essentially erase information from previous measurements.   As such, time reversal of the measurement process is established by reversing dynamics for a single measurement update step, where time reversed measurement `undoes' the backaction from forward measurement in a physically realizable way (Fig.~1b). This measurement reversal has been observed in a variety of experimental platforms \cite{katz06, katz08, kim11, murc13} and analyzed in the context of POVMs as we employ here \cite{para11}. For each measurement by POVM $M_{r_k}$, there is a corresponding measurement $\tilde{M}_{r_k}={M}_{\tilde{r}_k}$, where $\tilde{r}_k=-r_k$ is the time reversed measurement record which restores the initial state-of-knowledge, albeit with a statistical weight,
$$
\tilde{M}_{r_k}{M}_{r_k}\rho_k{M}_{r_k}^\dagger\tilde{M}_{r_k}^\dagger=\frac{\delta t}{2\pi\tau}{e^{-\frac{\delta t}{2\tau}(r_k^2+1)}}\rho_k.
$$
In addition, at each step the unitary evolution of the Rabi drive is reversed ($\Omega\to-\Omega$).}
To explore the statistical cost of time-reversed dynamics along a quantum trajectory with many timesteps, we use the path probability densities of forward and backward evolution to define a quantity $\lnR$ that characterizes the length of time's arrow,
\begin{equation}\label{eq:lnR}
\lnR=\sum_k \ln\frac{P(r_k|\rho_k)}{P(\tilde{r}_k|\rho_{k+1})}
=\sum_k\ln\frac{\tr{M_{r_k} \rho_k M_{r_k}^\dagger}\,dr_k}{\tr{\tilde{M}_{r_k} \rho_{k+1}\tilde{M}_{r_k}^\dagger}\,d\tilde{r}_k}.
\end{equation}
Since $\lnR$ is the sum of relative entropies between forward and reversed timesteps, a trajectory with positive $\lnR$ indicates a forward pointing arrow of time, for its measurement record has a greater probability density when considered forward, opposed to backward, in time. 

\redline{To gain an intuition into this arrow of time quantity $\lnR$ we examine the measurement record in continuous limit, $r(t) \propto z(t) + \sqrt{\tau} d \xi$, where $r(t)$ and $z(t)$ are respectively the measurement record and qubit expectation value $\langle\sigma_z\rangle$ in the continuous limit, and $d \xi$ is a zero mean Gaussian random variable. Consequently, we calculate Eq.~\ref{eq:lnR} as an integral and find $\dot{\lnR}_k\simeq2r(t)z(t)/\tau$ \cite{Dressel2017,SM}. Hence it is clear that the contributions to the forward arrow of time occur when the record and state are correlated.}

In our experiment, the measurements take place with a finite quantum efficiency $\eta = 0.4$\redline{, determined by comparing the measurement rate to the total dephasing rate $\Gamma = 1/(2\eta\tau)$} \cite{note}. Measurement with finite efficiency can be modeled by the measurement dynamics of multiple measurement channels, where our experimental measurement record is but one of these channels \cite{jaco06}. In the case of multiple channels, the dynamics is described by a POVM characterizing simultaneous measurement from every channel. An observer who has access to only one channel then describes system dynamics by averaging over all unknown measurement outcomes. Averaging over the unknown measurement outcomes results in dephasing of the qubit state, which breaks the time reversibility of the measurement dynamics.  To restore reversibility we estimate the quantum trajectories that would be obtained for an observer with access to all measurement channels. These trajectories serve as the model that governs the probability density for forward and reversed measurement sequences.

\begin{figure}\begin{center}
\includegraphics[angle = 0, width = .45\textwidth]{\FigPath 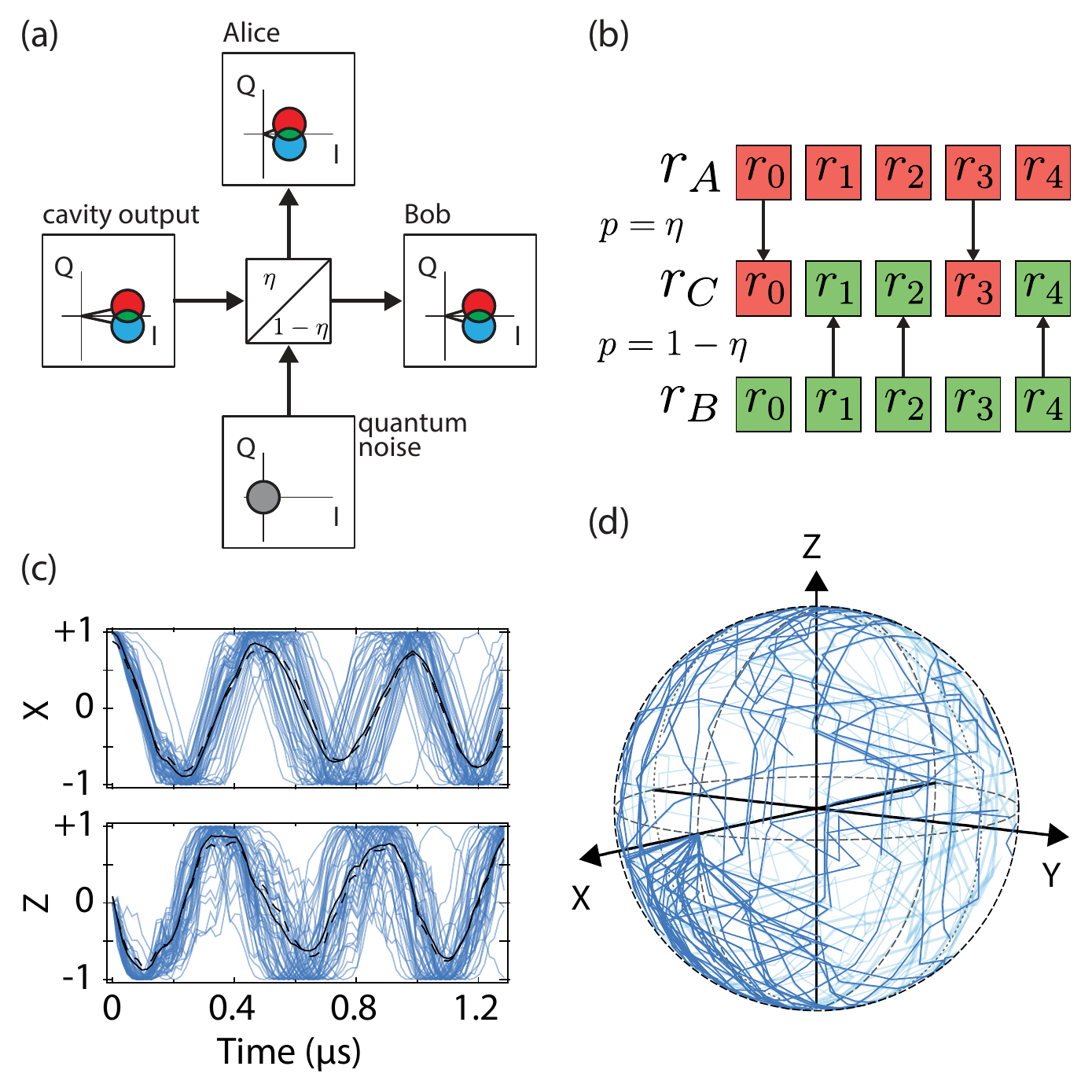}
\end{center}
\caption{Measurement inefficiency as multiple observers. (a) Finite quantum efficiency can be modeled as a beamsplitter, where the cavity probe is split between two observers, Alice and Bob. (b) We model the beamsplitter as a time segmented splitter, which directs the signal to Alice or Bob at each timestep with probabilities $\eta$ and $1-\eta$ respectively.  The measurement record of a third  observer, Charlie, who has access to both Alice and Bob's records can be constructed by taking either Alice's record or Bob's record at each timestep. (c) By sampling several possible measurement records for Bob, where Bob performs measurements on his signal in the same basis as Alice, we construct an ensemble of possible pure state trajectories for Charlie. The average of these unraveled trajectories (black solid line) , corresponds to the finite quantum efficiency trajectory based only on Alice's record (dashed line). (d) \redline{If Bob instead measures in a basis incompatible to Alice's measurement basis,} the resulting backaction causes evolution of the qubit outside the $X$--$Z$ plane. }
\end{figure}

As shown in Figure~2, the finite quantum efficiency in our experiment arises predominantly from attenuation of the cavity probe between the cavity and Josephson parametric amplifier, and this attenuation can be modeled as a beam splitter where the cavity probe is split between two observers which we denote ``Alice'' (our experimental record) and ``Bob'' (an unmonitored channel).  A third observer, ``Charlie'' has access to both Alice and Bob's measurement records and can therefore track quantum trajectories that are reversible as previously discussed.

\redline{For each experimentally sampled quantum trajectory we create an ensemble of pure state trajectories that are estimates for the pure state quantum trajectory determined by Charlie, as depicted schematically in Figure~2.} This ensemble of possible trajectories for Charlie corresponds to an unraveling of the \redline{Lindblad} master equation that describes Alice's quantum trajectory.  

This unraveling, however, depends on what type of measurements Bob makes on his channel, which we will analyze in the two extremal cases. \redline{Alice uses a parametric amplifier to measure the quadrature of the microwave probe that is correlated with the qubit populations in the $\sigma_z$ basis, which we denote as $z$-measurement. In one case, Bob measures the field in this basis revealing further information about the qubit populations.  In the other case, Bob measures in an incompatible basis, which does not reveal information about the qubit populations but rather about phase shifts on the qubit imparted by an ac Stark shift due to the cavity probe, which we refer to as $\phi$-measurement \cite{koro11,hatr13,murc13}. The corresponding quantum trajectories for Charlie are markedly different in both cases, resulting in different possible arrows of time for Alice.}

We model the finite efficiency beam splitter (Fig.~2a) in a time segmented fashion, such that Alice makes perfectly efficient measurements for a fraction $\eta$ of her measurement records, and records noise upon the remaining $1-\eta$ of measurement records. A possible trajectory for Charlie is constructed by updating the state with Alice's measurement record with probability $\eta$, otherwise with Bob's measurement record at each timestep. \redline{Both state updates are implemented with unit efficency ($1/\tau\to1/\eta\tau$).  In the two limiting cases for Bob's measurement basis, we construct Bob's measurement \redline{by statistically sampling measurement values distributed according to Charlie's state: Bob's compatible-basis-measurements are characterized by Eq.~1 and incompatible-basis-measurements are drawn from a zero-mean Gaussian of variance $1/(2\Gamma\delta t)=\eta\tau/\delta t$ \cite{SM}}. From a single sequence of experimentally obtained measurement records we create an ensemble of unraveled trajectories which has an average evolution consistent with the single finite efficiency experiment trajectory. Several unraveled quantum trajectories are shown for these two limiting cases in Figure 2c and 2d.}

\begin{figure}\begin{center}
\includegraphics[angle = 0, width = 0.48\textwidth]{\FigPath 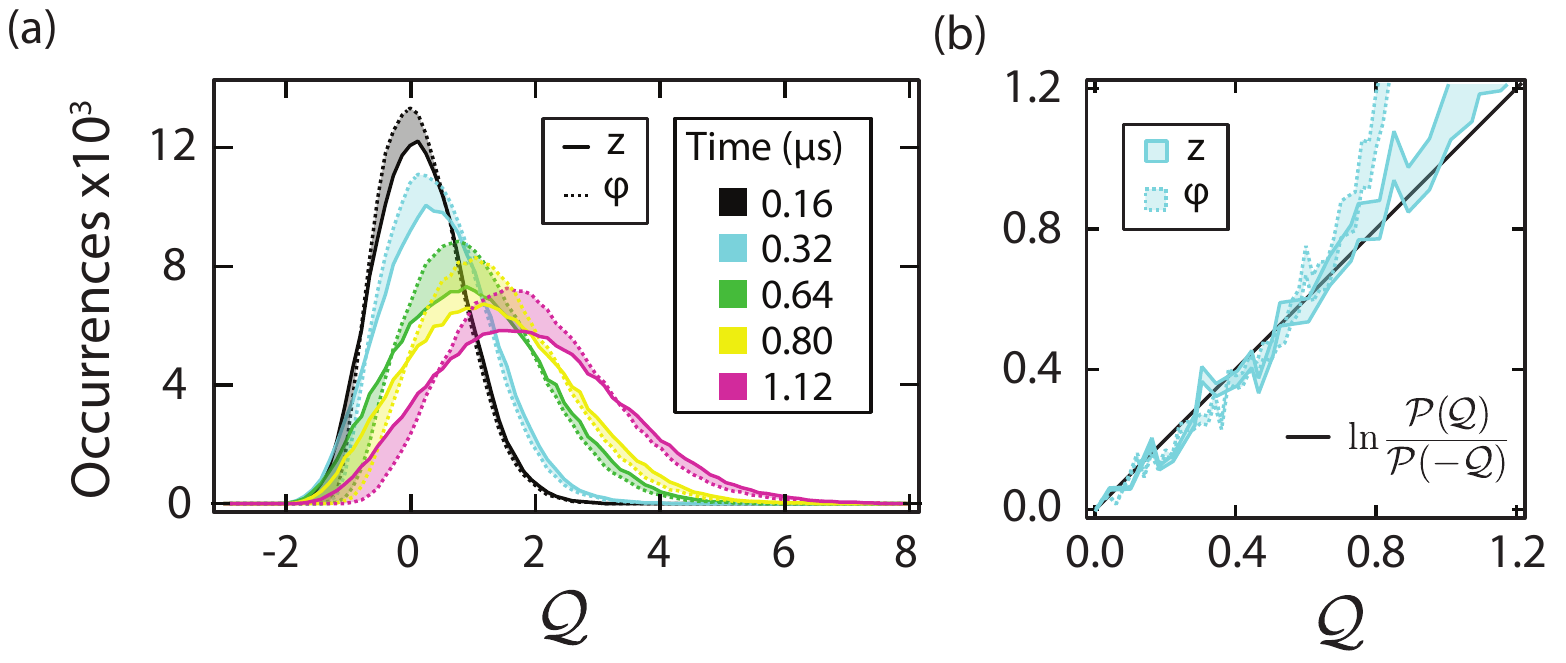}
\end{center}
\caption{Statistical arrow of time for quantum measurement evolution.  (a) The distributions $\mathcal{P}(\lnR)$ for different propagation times which are obtained from $2.8\times10^5$ runs of experiments. Here the solid curves correspond to Alice's arrow of time when Bob measures in the same basis as Alice, and the dashed curve indicates the case where Bob measures in the orthogonal basis, constraining a range of possible values for Alice's arrow of time. (b) Calculation of the detailed fluctuation theorem.  The distributions of $\lnR$ at time $t=0.32\ \mu$s (blue curves) is used to calculate the quantity $\ln (\mathcal{P}(\lnR)/\mathcal{P}(-\lnR))$. The detailed fluctuation theorem in Eq.(\ref{fluc}) for both of Bob's measurement bases agrees well with the theory prediction (black line). Error bars indicate the statistical uncertainty in quantity $\ln (\mathcal{P}(\lnR)/\mathcal{P}(-\lnR))$. }
\end{figure}

We now examine the arrow of time for an ensemble of experimentally sampled quantum trajectories. Figure~3a displays distributions of $\lnR$ based on $2.8\times10^5$ trajectories for different evolution times.  Each evolution time is associated with two different distributions for $\lnR$, corresponding to Alice's arrow of time, given the limiting cases of Bob's measurement basis. \redline{Here we see the role of measurement backaction in the choice of Bob's measurement basis, where $z$-measurement leads to a greater occurrence of both forward-likely and backward-likely trajectories as indicated by the broad $\lnR$ distribution compared to the counterpart $\lnR$ distribution for $\phi$-measurements. When Bob measures in a compatible basis with Alice, the effectively stronger measurement causes the trajectory to take on more extremal values of $z$ resulting in stronger correlation and anti-correlation to the observed measurement record.}

Notably, negative values of $\lnR$ occur for Alice's arrow of time \redline{for both Bob's $z$ and $\phi$-measurements,} corresponding to trajectories where the time reverse process is more likely.  This phenomenon of negative entropy production is well known in microscopic stochastic systems and is typically characterized by a fluctuation theorem \cite{seif05,croo00,harr07,horo11,croo98,jarz99,espo10,jar11,seif12,cil13}.  In Figure~3b we show that the data are in agreement with a detailed fluctuation theorem,
\begin{equation}\label{fluc}
\frac{\mathcal{P}(\lnR)}{\mathcal{P}(-\lnR)}=e^\lnR, 
\end{equation}
which quantifies the relative probability of obtaining a forward pointing arrow of time with length $\lnR$ to the probability of a backwards arrow of the same length. For small values of $\lnR$, the agreement with the detailed fluctuation theorem indicates that the experimentally sampled relative occurrence of $\lnR$, as given by the left hand side of Eq.~\ref{fluc}, is in agreement with the definition of $\lnR$ on the right hand side. The self consistency of such a fluctuation arises because the microscopic dynamics and the macroscopic statistics are consistent with the definition of $\lnR$. \redline{However, for larger values of $\lnR$, the fluctuation relation is clearly nonlinear, a feature that is related to the presence of absolute irreversibility which we now explore in detail.}

\redline{To examine the presence of absolute irreversibility, we consider special measurement condition where the Rabi frequency of the drive $\Omega = 0$}.  In this case, the measurement operators commute with the qubit Hamiltonian resulting in a quantum non-demolition measurement.  We consider the case where the qubit is prepared such that $\langle \sigma_x\rangle \simeq 1$ and measurements project the system toward the stationary points $\langle\sigma_z\rangle \rightarrow \pm1$.  Figure~4 displays the distributions for $\lnR$ for several evolution times. Note that Bob's measurement basis does not affect Alice's arrow of time in this case.

For the simple dynamics of this semi-classical measurement, the probability density of $\lnR$ is found analytically by performing a change of variables in the measurement record probability density \cite{Dressel2017},
\begin{equation}\label{eq:distQ}
\mathcal{P}(\lnR)=\sqrt{\frac{T}{2\pi\tau}\frac{e^{\lnR}}{e^{\lnR}-1}}\exp\!\left\{\!-\frac{T}{2\tau}-\frac{\tau}{2T}[\mathrm{cosh}^{-1}(e^{\lnR/2})]^2\right\}.
\end{equation}
Histograms of $\lnR$ from experiment are plotted for a selection of final times $T$ with their corresponding theoretical probability density in dashed lines.

Clearly, the relative probabilities for forward and backwards arrows of time in this measurement case do not satisfy the detailed fluctuation theorem (Eq.~\ref{fluc}). This is because the detailed fluctuation relation is only satisfied for the total statistical entropy change during a process \cite{seif05}. In the presented case, the statistical arrow of time quantity $\lnR$ does not capture the contributing influence of the initial state of the trajectory, hence quantum measurement is, in general, a nonequilibrium, irreversible process. Here, the initial state imposes a lower bound on the possible values of $\lnR$ \cite{Dressel2017}. This sensitivity to initial conditions results from the `un'-likelihood of a particular initial state, quantified by an absolute irreversibility \cite{mura14, funo15, manz15,sree19}. As presented in Figure~4b, the absolute irreversibility is quantified by the integral fluctuation theorem, which gives a deviation from unity resulting from the ensemble of trajectories containing a surplus of state updates that have a positive statistical arrow of time. This is due to the favoring of correlations between the qubit state and measure record from the measurement projection process. This contribution to the entropy is physically analogous to the entropy increase associated with irreversible expansion of gas. The semi-classical measurement case discussed here clearly illustrates absolute irreversibility due to initial conditions since the initial state is far from the fixed points of the measurement dynamics. 

\begin{figure}\begin{center}
\includegraphics[angle = 0, width = .48\textwidth]{\FigPath 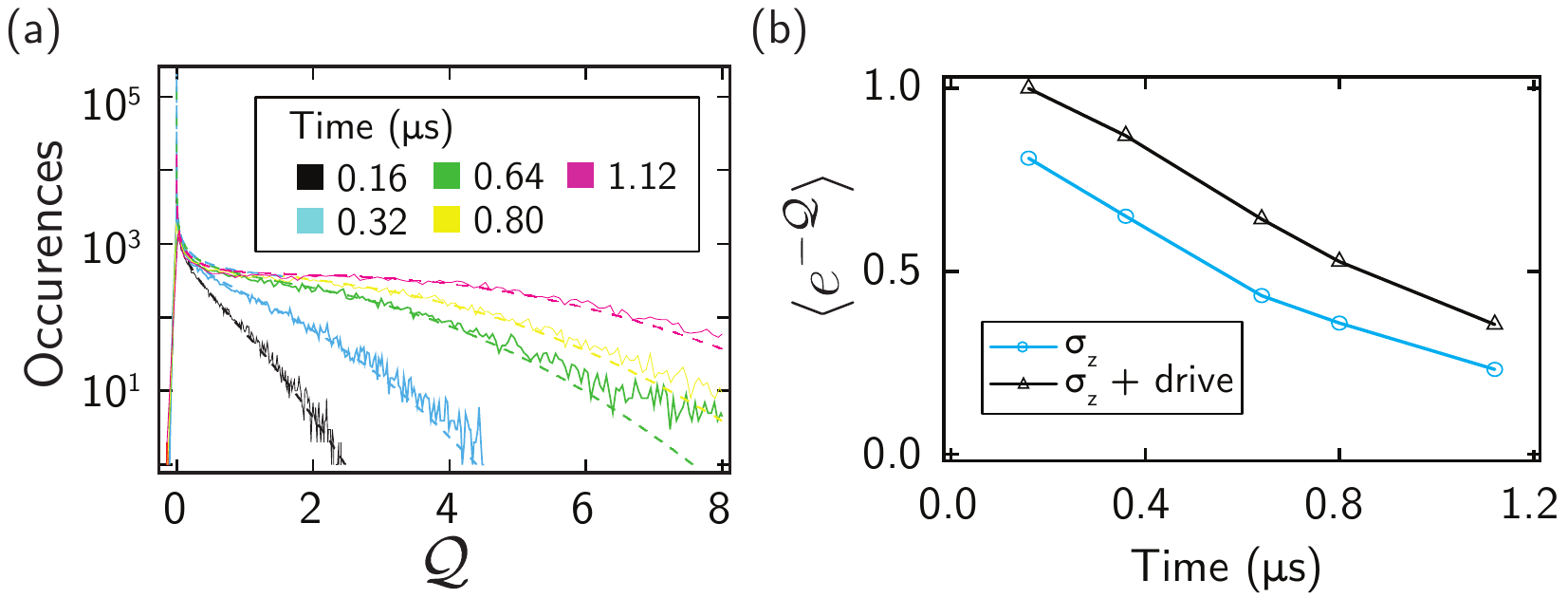}
\end{center}
\caption{\small{Distribution of $\lnR$ and absolute irreversibility. (a) The distribution of $\lnR$ for different propagation times obtained from $2.8\times10^5$ experimental trajectories.  As the time increases, the distribution is more biased to large positive values of $\lnR$.  (b) Calculation of the absolute irreversibility in the case Bob measures in the same basis as Alice. The integral fluctuations theorem gives a quantity less than unity as a consequence of the initial state of the trajectory. Here, the dynamics of state projection favor measurement records correlated with the qubit state, resulting in a surplus of state updates when $\mathcal{Q}>0$, causing an overall longer arrow of time.}}
\end{figure} 

Continuous quantum measurement leads to a probabilistic dynamics of the quantum state. Under conditions where this dynamics is time reversible, we can consider the probabilities associated with both forward and reversed dynamics.  These probabilities allow us to infer a statistical arrow of time by considering the information entropy associated to measurement sequences.  Using experimental data we have shown that a statistical arrow of time emerges fundamentally in quantum measurement, where information and backaction arise from entanglement with a fluctuating environment.

\begin{acknowledgements}
\emph{Acknowledgements}---We acknowledge S. Weber and I. Siddiqi for data contributions and A. Chantasri, S. K. Manikandan, and A. Jordan for extensive guidance and K. Jacobs for fruitful discussions. This research was supported in part by the John Templeton Foundation Grant ID 58558, the Sloan Foundation, and used facilities at the Institute of Materials Science and Engineering at Washington University.  D.T. acknowledges support from the Rigetti Computing Postdoctoral Fellowship.
\end{acknowledgements}

\noindent * Denotes equal contribution.\\
$^\dagger$ patrick.harrington@wustl.edu, murch@wustl.edu
%

\onecolumngrid
\section*{Supplementary Material}

This supplementary material contains further information of the arrow of time expressions in the presence of finite measurement efficiency as well as further details on the unraveling procedure presented in the main text.

\section{Quantum state tracking with finite efficiency measurements}
As sketched in Figure 1, the finite efficiency measurements can be modeled as multiple observers that each measure a portion of the cavity output. While in the main text, we considered only two observers, Alice and Bob, where Bob has a choice to measure his observed signal in any basis, here we consider three observers; Alice who measures a portion of the signal in the basis that corresponds to information about the qubit populations in the energy basis ($z$-measurement), an additional observer Bob, who also measures in this basis, and a third observer, Rob, who measures in an incompatible basis ($\phi$-measurement).  We also consider a global observer, Charlie, who has access to all the measurement records (and can further exclude the added quantum noise from the beam splitter).

 \subsection{Multiple measurement records: Alice ($z$), Bob ($z$), \& Rob ($\phi$)}
\begin{figure}
\begin{center}
\includegraphics[width = 0.8\textwidth]{\FigPath 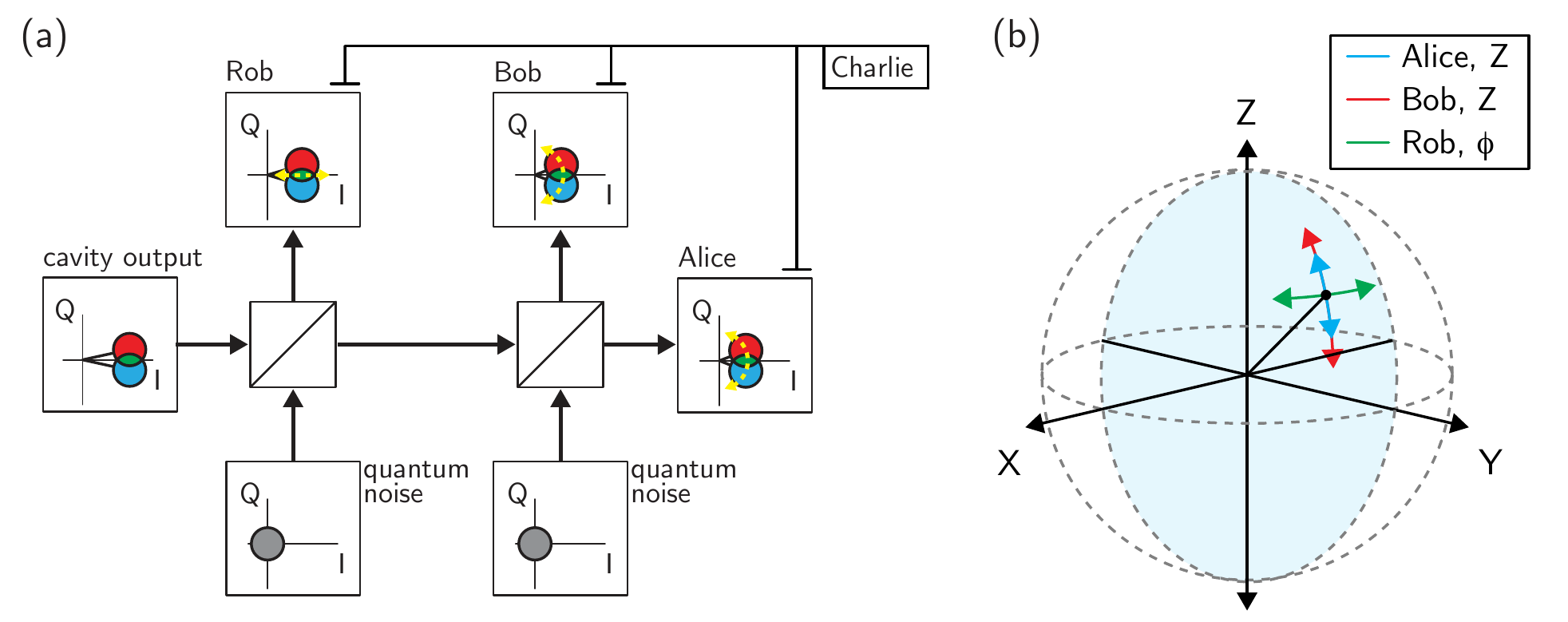}
\end{center}
\caption{\small{Multiple measurement channels: Alice, Bob, and Rob.} (a) We use a beamsplitter model to describe the decoherence effects due to measurement inefficiency.  Alice and Bob measure the cavity output in compatible bases corresponding to $z$-measurements; measurements of the qubit-state-dependent phase shift of the cavity output.  Rob measures the cavity output in an incompatible basis, measuring instead the cavity output photon number, denoted $\phi$-measurement. A third observer, Charlie, has access to all the measurement records. (b) Each observer's measurement induces backaction on Charlie's pure quantum state.}\label{fig5}
\end{figure}

We begin our discussion by constructing a positive operator-valued measure (POVM) for Charlie's measurement given all measurement records from Alice, Bob, and Rob which we denote respectively as $r$, $\vartheta_z$, and $\vartheta_\phi$.  This POVM is constructed from three commuting measurements, 
\begin{align*}
M_{r,\vartheta_z,\vartheta_\phi}&=M_{\vartheta_\phi}M_{\vartheta_z}M_{r_R}\\
&=
\bigg(\frac{\delta t}{{2\pi}}\bigg)^{3/4}\bigg(\frac{4\gamma_z\gamma_\phi}{\tau}\bigg)^{1/4}\! \times
\\
&\hspace{1cm}
\exp\!\bigg[\!
-\frac{\delta t}{4\tau}(r-\sigma_z)^2
-\frac{\gamma_z\delta t}{2}(\vartheta_z-\sigma_z)^2
-\frac{\gamma_\phi\delta t}{2}(\vartheta_\phi^2+2i\vartheta_\phi\sigma_z)
\bigg],
\end{align*}
where $1/\tau$, $2\gamma_z$, and $2\gamma_\phi$ are the measurement rates for Alice, Bob, and Rob respectively. Charlie can reconstruct the qubit state evolution informed by both the experimentally measured signal from Alice and the experimentally inaccessible measurement records from Bob and Rob following the usual update equations for the density operator,
$$
\rho_{k+1} = \frac{M_{r,\vartheta_z,\vartheta_\phi} \rho_k M_{r,\vartheta_z,\vartheta_\phi}^\dagger}{\mathrm{tr}[M_{r,\vartheta_z,\vartheta_\phi} \rho_k M_{r,\vartheta_z,\vartheta_\phi}^\dagger]}.
$$ 
Furthermore, if Charlie averages over all measurement records, he observes an ensemble dephasing rate with contributions from each measurement channel,
\begin{equation}\label{eq:dephase}
\Gamma=
\frac{1}{2\tau}+\underbrace{\frac{1}{2\tau_z}}_{\gamma_z}+\underbrace{\frac{1}{2\tau_\phi}}_{\gamma_\phi}
=\frac{1}{2\tau}+{\gamma_z}+{\gamma_\phi},
\end{equation}
where $1/2\tau$, $\gamma_z$, and $\gamma_\phi$ are respectively the dephasing rates induced by Alice, Bob, and Rob's measurement channels. Note that every observer agrees on the total ensemble dephasing rate $\Gamma$, for this dynamics results from ignoring all measurement records. 

\subsection{Dephasing and the POVM for Alice's measurement}
With a full description of the measurement process from Charlie's POVM we are in a position to examine how each  measurement observer accounts for missing information of the other measurement channels, since Alice, Bob, nor Rob have a POVM that correctly describes the ensemble statistics and dephasing. Hence, Alice, Bob, and Rob each update their qubit state-of-knowledge using Charlie's POVM but must average over all possible measurement records from the other observers.

Thus Alice's updated state, $\rho_{k+1}^\mathrm{A}$, is
\begin{equation}\label{eq:dephasing0}
\rho_{k+1}^\mathrm{A} \propto \int d\vartheta_zd\vartheta_\phi\,
{M_{r,\vartheta_z,\vartheta_\phi}\rho_{k}^\mathrm{A} M_{r,\vartheta_z,\vartheta_\phi}^{\dagger}}.
\end{equation}
In terms of state update equations, this expression is equivalent to an update with Alice's POVM and then performing a dephasing operation by rescaling off-diagonal elements. This POVM is written as \cite{jord06}, 
\begin{equation}
M_{r}=
\bigg({\frac{{\delta t}}{{2\pi\tau}}}\bigg)^{1/4}\!
\exp\!\bigg[\!-\frac{\delta t}{4\tau}(r-\sigma_z)^2\bigg], \label{eq:qnd}
\end{equation}
where Alice's POVM update is followed by a dephasing operation which rescales off-diagonal density matrix elements by a factor $e^{-(\Gamma-1/2\tau)\delta t}=e^{-(\gamma_z+\gamma_\phi)\delta t}$. 

For the case of measurement in the presence of an additional Rabi drive with Rabi frequency $\Omega$, each state update step can be approximated as a measurement given by (\ref{eq:qnd}) followed by unitary rotation by angle $\Omega\delta t$ in the Bloch sphere on the axis of the drive. We find measurement updates from this POVM are consistent with the qubit state evolution as verified by quantum state tomography \cite{murc13, webe14}.

\subsection{Forward \& backward path probabilities}
The notion of time reversal of the measurement presented in the main text is only valid for the perfect efficiency observation characterized by Charlie's POVM measurement, since inefficient measurement does not lend to time reversible dynamics due to pure dephasing. Here, we examine the arrow of time statistic for Charlie's observation and show that it is a sum of Alice's arrow of time statistic and Bob's statistic, thereby justifying the expression used in the main text.

In this calculation, we first consider the path probability density for continuous weak measurement trajectories of Charlie's POVM. We then calculate the arrow of time statistic for Charlie's trajectories $\lnR_c$ as the log ratio of Charlie's forward and backward path probabilities according to the procedure of \cite{Dressel2017}.

The joint probability distribution for Alice, Bob, and Rob's measurement outcomes ($r$, $\vartheta_z$, and $\vartheta_\phi$) given the qubit state $\rho$ is calculated from Charlie's POVM as,
\begin{align}
\begin{split}\label{eq:pathprob}
P_{\delta t}(r,\vartheta_z,\vartheta_\phi)\,drd\vartheta_zd\vartheta_\phi
&=
\mathrm{tr}[M_{r,\vartheta_z,\vartheta_\phi}\rho M_{r,\vartheta_z,\vartheta_\phi}^{\dagger}]
\,drd\vartheta_zd\vartheta_\phi\\
&=
\bigg(\frac{\delta t}{{\pi}}\bigg)^{3/2}\!\sqrt{\frac{\gamma_z\gamma_\phi}{2\tau}}
\exp\!\big(\!-\gamma_\phi\delta t\,\vartheta_\phi^2\big)  \times \\
& \hspace{2cm} \bigg(\frac{1+z}{2}\exp\!\bigg[\!-\frac{\delta t}{2\tau}(r-1)^2-\gamma_z\delta t(\vartheta_z-1)^2\bigg]+
\\
&\hspace{3cm}
\frac{1-z}{2}\exp\!\bigg[\!-\frac{\delta t}{2\tau}(r+1)^2-\gamma_z\delta t(\vartheta_z+1)^2\bigg]
\bigg)\,drd\vartheta_zd\vartheta_\phi.
\end{split}
\end{align}
We rewrite Eq.~(\ref{eq:pathprob}) in the continuous limit ($\delta t\ll\tau,1/2\gamma_z,1/2\gamma_\phi$) and only consider exponentiated terms since we are interested in path probability ratios,
\begin{align*}
P_{\delta t}(r,\vartheta_z,\vartheta_\phi)\,drd\vartheta_zd\vartheta_\phi &\propto
\exp\!\bigg[
-\frac{\delta t}{2\tau}(r^2-2rz+1)
-{\gamma_z\delta t}\,(\vartheta_z^2-2\vartheta_z z+1)
-{\gamma_\phi\delta t}\,\vartheta_\phi^2
\bigg]
\,drd\vartheta_zd\vartheta_\phi.
\end{align*}
A given continuous trajectory of duration $T$ then samples from the entire path distribution that is the continuous product of state updates,
\begin{align*}
P(r(t),\vartheta_z(t),\vartheta_\phi(t))
&\,\mathcal{D}r\mathcal{D}\vartheta_z\mathcal{D}\vartheta_\phi=
\exp\!\bigg[\int_0^T dt\,
\bigg(\!-\frac{r(t)^2+1}{2\tau}-\gamma_z(\vartheta_z(t)^2+1)-\gamma_\phi\vartheta_\phi(t)^2\bigg)
+\\
&\hspace{5.5cm}+\bigg(\frac{r(t)}{\tau}+2\gamma_z\vartheta_z(t)\bigg)z(t)\bigg]
\,\mathcal{D}r\mathcal{D}\vartheta_z\mathcal{D}\vartheta_\phi,
\end{align*}
where exponential prefactors are absorbed by the functional measures $\mathcal{D}r\mathcal{D}\vartheta_z\mathcal{D}\vartheta_\phi$. 

To find the relative path probability between forward-in-time and backward-in-time trajectories we apply a time reversing transformation to the probability distribution by the replacement $t\to T-t$ and flipping the sign of the measurement records
$$
\begin{array}{ccc}
r(t)\to -r(T-t), & \vartheta_z(t)\to -\vartheta_z(T-t), &\vartheta_\phi(t)\to -\vartheta_\phi(T-t),
\end{array}
$$
such that the time reversed trajectories adhere to the same equations of motion of their forward-in-time counterparts, ensuring reversible dynamics. The time reversed probability density is,
\begin{align*}
\tilde{P}(r(t),\vartheta_z(t),\vartheta_\phi(t))
&\,\mathcal{D}r\mathcal{D}\vartheta_z\mathcal{D}\vartheta_\phi=
\exp\!\bigg[-\int_0^T dt\,
\bigg(\!-\frac{r(T-t)^2+1}{2\tau}-\gamma_z(\vartheta_z(T-t)^2+1)-\gamma_\phi\vartheta_\phi(T-t)^2\bigg)
+\\
&\hspace{5.5cm}-\bigg(\frac{r(T-t)}{\tau}+2\gamma_z\vartheta_z(T-t)\bigg)z(T-t)\bigg]
\,\mathcal{D}r\mathcal{D}\vartheta_z\mathcal{D}\vartheta_\phi.
\end{align*}
We calculate the arrow of time statistic as the logarithm ratio of the forward and backward probability densities,
\begin{align}\label{eq:Qcharlie}
\begin{split}
\lnR_\mathrm{C}&=\ln\frac{{P}(r(t),\vartheta_z(t),\vartheta_\phi(t))}{\tilde{P}(r(t),\vartheta_z(t),\vartheta_\phi(t))}\\
&=
2\int_0^T dt\,\bigg(\underbrace{\frac{r(t)}{\tau}z(t)}_{\text{Alice}}+\underbrace{2\gamma_z\vartheta_z(t)z(t)}_{\text{Bob}}\bigg),
\end{split}
\end{align}
and we note two points from  Eq.~\ref{eq:Qcharlie}. First, Charlie's arrow-of-time statistic does not depend on Rob's measurements, since Rob's measurement record shares no correlation with the measured state, thus Rob's measurement term is statistically invariant under time reversal. Second, we note that $\lnR_\mathrm{C}$ is a sum of two terms; the first term is the correlation between Alice's record, $r(t)$ and Charlie's state $z(t)$, the second is the correlation between Bob's record $\vartheta_z(t)$ and Charlie's state. In the main text, we presented Alice's observed arrow-of-time, which is given by the first term of Eq.~\ref{eq:Qcharlie}. The effect of Bob and Rob's measurements on Alice's arrow-of-time enters through Charlie's state $z(t)$.

In the case of measurement along with Rabi drive, for the weak driving limit $\Omega \delta t\ll1$, we simply apply this reasoning at each timestep, summing the contributions of each measurement to Alice's arrow-of-time.

\section{Unravelling method for inefficient experiment trajectories}

For the unravelling of experimental weak measurement trajectories, we replace the beamsplitter discussion presented above with a method that is well suited for the piecewise continuous trajectories that we study. Here we model the beamsplitter in a time segmented fashion, where at any timestep the measurement signal is directed to one of the observers Alice, Bob, or Rob with unit efficiency and the other two observers measure zero-mean Gaussian noise. Charlie in turn knows the recorded values and the configuration of the beamsplitters at each timestep.  In this approach, a measurement channel is probabilistically selected to perform a measurement for a single timestep, measuring with the total measurement strength $2\Gamma$ set by the ensemble dephasing rate $\Gamma$ as defined in Eq.~(\ref{eq:dephase}). 

The measurement strengths $1/2\tau$, $2\gamma_z$, and $2\gamma_\phi$ fix the relative likelihood that a measurement channel is randomly selected at a given time. Here, inefficient experimental measurement is equivalent to Alice partaking in qubit measurement for only a fraction of all time, but with perfectly efficient measurements and a measurement strength according to the total dephasing rate.

This method makes construction of an ensemble of possible trajectories for Charlie possible; for each of Alice's records, a random sample of $\eta$ of these records are chosen as real measurements, updated with Eq.~\ref{eq:qnd} using $\tau \rightarrow 1/2\Gamma=\eta\tau$. For the other measurement steps, we consider hypothetical measurements given by either Bob or Rob corresponding to the two limiting cases where they each measure the remaining signal with perfect efficiency.

\bibliography{references}

\begin{thebibliography}{90}%
\makeatletter
\providecommand \@ifxundefined [1]{%
 \@ifx{#1\undefined}
}%
\providecommand \@ifnum [1]{%
 \ifnum #1\expandafter \@firstoftwo
 \else \expandafter \@secondoftwo
 \fi
}%
\providecommand \@ifx [1]{%
 \ifx #1\expandafter \@firstoftwo
 \else \expandafter \@secondoftwo
 \fi
}%
\providecommand \natexlab [1]{#1}%
\providecommand \enquote  [1]{``#1''}%
\providecommand \bibnamefont  [1]{#1}%
\providecommand \bibfnamefont [1]{#1}%
\providecommand \citenamefont [1]{#1}%
\providecommand \href@noop [0]{\@secondoftwo}%
\providecommand \href [0]{\begingroup \@sanitize@url \@href}%
\providecommand \@href[1]{\@@startlink{#1}\@@href}%
\providecommand \@@href[1]{\endgroup#1\@@endlink}%
\providecommand \@sanitize@url [0]{\catcode `\\12\catcode `\$12\catcode
  `\&12\catcode `\#12\catcode `\^12\catcode `\_12\catcode `\%12\relax}%
\providecommand \@@startlink[1]{}%
\providecommand \@@endlink[0]{}%
\providecommand \url  [0]{\begingroup\@sanitize@url \@url }%
\providecommand \@url [1]{\endgroup\@href {#1}{\urlprefix }}%
\providecommand \urlprefix  [0]{URL }%
\providecommand \Eprint [0]{\href }%
\providecommand \doibase [0]{http://dx.doi.org/}%
\providecommand \selectlanguage [0]{\@gobble}%
\providecommand \bibinfo  [0]{\@secondoftwo}%
\providecommand \bibfield  [0]{\@secondoftwo}%
\providecommand \translation [1]{[#1]}%
\providecommand \BibitemOpen [0]{}%
\providecommand \bibitemStop [0]{}%
\providecommand \bibitemNoStop [0]{.\EOS\space}%
\providecommand \EOS [0]{\spacefactor3000\relax}%
\providecommand \BibitemShut  [1]{\csname bibitem#1\endcsname}%
\let\auto@bib@innerbib\@empty
\bibitem [{\citenamefont {Braginsky}\ and\ \citenamefont
  {Khalili}(1992)}]{bragbook}%
  \BibitemOpen
  \bibfield  {author} {\bibinfo {author} {\bibfnamefont {V.~B.}\ \bibnamefont
  {Braginsky}}\ and\ \bibinfo {author} {\bibfnamefont {F.~Y.}\ \bibnamefont
  {Khalili}},\ }\href@noop {} {\emph {\bibinfo {title} {Quantum Measurement}}}\
  (\bibinfo  {publisher} {Cambridge University Press},\ \bibinfo {year}
  {1992})\BibitemShut {NoStop}%
\bibitem [{\citenamefont {Wiseman}\ and\ \citenamefont
  {Milburn}(2010)}]{wisebook}%
  \BibitemOpen
  \bibfield  {author} {\bibinfo {author} {\bibfnamefont {H.}~\bibnamefont
  {Wiseman}}\ and\ \bibinfo {author} {\bibfnamefont {G.}~\bibnamefont
  {Milburn}},\ }\href@noop {} {\emph {\bibinfo {title} {Quantum Measurement and
  Control}}}\ (\bibinfo  {publisher} {Cambridge University Press},\ \bibinfo
  {year} {2010})\BibitemShut {NoStop}%
\bibitem [{\citenamefont {Jacobs}(2014)}]{jacobsbook}%
  \BibitemOpen
  \bibfield  {author} {\bibinfo {author} {\bibfnamefont {K.}~\bibnamefont
  {Jacobs}},\ }\href@noop {} {\emph {\bibinfo {title} {Quantum Measurement
  Theory}}}\ (\bibinfo  {publisher} {Cambridge},\ \bibinfo {year}
  {2014})\BibitemShut {NoStop}%
\bibitem [{\citenamefont {Guerlin}\ \emph {et~al.}(2007)\citenamefont
  {Guerlin}, \citenamefont {Bernu}, \citenamefont {Deleglise}, \citenamefont
  {Sayrin}, \citenamefont {Gleyzes}, \citenamefont {Kuhr}, \citenamefont
  {Brune}, \citenamefont {Raimond},\ and\ \citenamefont {Haroche}}]{guer07}%
  \BibitemOpen
  \bibfield  {author} {\bibinfo {author} {\bibfnamefont {C.}~\bibnamefont
  {Guerlin}}, \bibinfo {author} {\bibfnamefont {J.}~\bibnamefont {Bernu}},
  \bibinfo {author} {\bibfnamefont {S.}~\bibnamefont {Deleglise}}, \bibinfo
  {author} {\bibfnamefont {C.}~\bibnamefont {Sayrin}}, \bibinfo {author}
  {\bibfnamefont {S.}~\bibnamefont {Gleyzes}}, \bibinfo {author} {\bibfnamefont
  {S.}~\bibnamefont {Kuhr}}, \bibinfo {author} {\bibfnamefont {M.}~\bibnamefont
  {Brune}}, \bibinfo {author} {\bibfnamefont {J.}~\bibnamefont {Raimond}}, \
  and\ \bibinfo {author} {\bibfnamefont {S.}~\bibnamefont {Haroche}},\
  }\href@noop {} {\bibfield  {journal} {\bibinfo  {journal} {Nature}\ }\textbf
  {\bibinfo {volume} {448}},\ \bibinfo {pages} {889} (\bibinfo {year}
  {2007})}\BibitemShut {NoStop}%
\bibitem [{\citenamefont {Murch}\ \emph {et~al.}(2013)\citenamefont {Murch},
  \citenamefont {Weber}, \citenamefont {Macklin},\ and\ \citenamefont
  {Siddiqi}}]{murc13}%
  \BibitemOpen
  \bibfield  {author} {\bibinfo {author} {\bibfnamefont {K.~W.}\ \bibnamefont
  {Murch}}, \bibinfo {author} {\bibfnamefont {S.~J.}\ \bibnamefont {Weber}},
  \bibinfo {author} {\bibfnamefont {C.}~\bibnamefont {Macklin}}, \ and\
  \bibinfo {author} {\bibfnamefont {I.}~\bibnamefont {Siddiqi}},\ }\href@noop
  {} {\bibfield  {journal} {\bibinfo  {journal} {Nature}\ }\textbf {\bibinfo
  {volume} {502}},\ \bibinfo {pages} {211} (\bibinfo {year}
  {2013})}\BibitemShut {NoStop}%
\bibitem [{\citenamefont {Roch}\ \emph {et~al.}(2014)\citenamefont {Roch},
  \citenamefont {Schwartz}, \citenamefont {Motzoi}, \citenamefont {Macklin},
  \citenamefont {Vijay}, \citenamefont {Eddins}, \citenamefont {Korotkov},
  \citenamefont {Whaley}, \citenamefont {Sarovar},\ and\ \citenamefont
  {Siddiqi}}]{roch14}%
  \BibitemOpen
  \bibfield  {author} {\bibinfo {author} {\bibfnamefont {N.}~\bibnamefont
  {Roch}}, \bibinfo {author} {\bibfnamefont {M.~E.}\ \bibnamefont {Schwartz}},
  \bibinfo {author} {\bibfnamefont {F.}~\bibnamefont {Motzoi}}, \bibinfo
  {author} {\bibfnamefont {C.}~\bibnamefont {Macklin}}, \bibinfo {author}
  {\bibfnamefont {R.}~\bibnamefont {Vijay}}, \bibinfo {author} {\bibfnamefont
  {A.~W.}\ \bibnamefont {Eddins}}, \bibinfo {author} {\bibfnamefont
  {A.}~\bibnamefont {Korotkov}}, \bibinfo {author} {\bibfnamefont {K.~B.}\
  \bibnamefont {Whaley}}, \bibinfo {author} {\bibfnamefont {M.}~\bibnamefont
  {Sarovar}}, \ and\ \bibinfo {author} {\bibfnamefont {I.}~\bibnamefont
  {Siddiqi}},\ }\href {\doibase 10.1103/PhysRevLett.112.170501} {\bibfield
  {journal} {\bibinfo  {journal} {Phys. Rev. Lett.}\ }\textbf {\bibinfo
  {volume} {112}},\ \bibinfo {pages} {170501} (\bibinfo {year}
  {2014})}\BibitemShut {NoStop}%
\bibitem [{\citenamefont {Campagne-Ibarcq}\ \emph {et~al.}(2016)\citenamefont
  {Campagne-Ibarcq}, \citenamefont {Six}, \citenamefont {Bretheau},
  \citenamefont {Sarlette}, \citenamefont {Mirrahimi}, \citenamefont
  {Rouchon},\ and\ \citenamefont {Huard}}]{camp16}%
  \BibitemOpen
  \bibfield  {author} {\bibinfo {author} {\bibfnamefont {P.}~\bibnamefont
  {Campagne-Ibarcq}}, \bibinfo {author} {\bibfnamefont {P.}~\bibnamefont
  {Six}}, \bibinfo {author} {\bibfnamefont {L.}~\bibnamefont {Bretheau}},
  \bibinfo {author} {\bibfnamefont {A.}~\bibnamefont {Sarlette}}, \bibinfo
  {author} {\bibfnamefont {M.}~\bibnamefont {Mirrahimi}}, \bibinfo {author}
  {\bibfnamefont {P.}~\bibnamefont {Rouchon}}, \ and\ \bibinfo {author}
  {\bibfnamefont {B.}~\bibnamefont {Huard}},\ }\href {\doibase
  10.1103/PhysRevX.6.011002} {\bibfield  {journal} {\bibinfo  {journal} {Phys.
  Rev. X}\ }\textbf {\bibinfo {volume} {6}},\ \bibinfo {pages} {011002}
  (\bibinfo {year} {2016})}\BibitemShut {NoStop}%
\bibitem [{\citenamefont {Naghiloo}\ \emph {et~al.}(2016)\citenamefont
  {Naghiloo}, \citenamefont {Foroozani}, \citenamefont {Tan}, \citenamefont
  {Jadbabaie},\ and\ \citenamefont {Murch}}]{nagh16}%
  \BibitemOpen
  \bibfield  {author} {\bibinfo {author} {\bibfnamefont {M.}~\bibnamefont
  {Naghiloo}}, \bibinfo {author} {\bibfnamefont {N.}~\bibnamefont {Foroozani}},
  \bibinfo {author} {\bibfnamefont {D.}~\bibnamefont {Tan}}, \bibinfo {author}
  {\bibfnamefont {A.}~\bibnamefont {Jadbabaie}}, \ and\ \bibinfo {author}
  {\bibfnamefont {K.~W.}\ \bibnamefont {Murch}},\ }\href@noop {} {\bibfield
  {journal} {\bibinfo  {journal} {Nature Communications}\ }\textbf {\bibinfo
  {volume} {7}},\ \bibinfo {pages} {11527} (\bibinfo {year}
  {2016})}\BibitemShut {NoStop}%
\bibitem [{\citenamefont {Chantasri}\ \emph {et~al.}(2013)\citenamefont
  {Chantasri}, \citenamefont {Dressel},\ and\ \citenamefont {Jordan}}]{chan13}%
  \BibitemOpen
  \bibfield  {author} {\bibinfo {author} {\bibfnamefont {A.}~\bibnamefont
  {Chantasri}}, \bibinfo {author} {\bibfnamefont {J.}~\bibnamefont {Dressel}},
  \ and\ \bibinfo {author} {\bibfnamefont {A.~N.}\ \bibnamefont {Jordan}},\
  }\href {\doibase 10.1103/PhysRevA.88.042110} {\bibfield  {journal} {\bibinfo
  {journal} {Phys. Rev. A}\ }\textbf {\bibinfo {volume} {88}},\ \bibinfo
  {pages} {042110} (\bibinfo {year} {2013})}\BibitemShut {NoStop}%
\bibitem [{\citenamefont {Weber}\ \emph {et~al.}(2014)\citenamefont {Weber},
  \citenamefont {Chantasri}, \citenamefont {Dressel}, \citenamefont {Jordan},
  \citenamefont {Murch},\ and\ \citenamefont {Siddiqi}}]{webe14}%
  \BibitemOpen
  \bibfield  {author} {\bibinfo {author} {\bibfnamefont {S.~J.}\ \bibnamefont
  {Weber}}, \bibinfo {author} {\bibfnamefont {A.}~\bibnamefont {Chantasri}},
  \bibinfo {author} {\bibfnamefont {J.}~\bibnamefont {Dressel}}, \bibinfo
  {author} {\bibfnamefont {A.~N.}\ \bibnamefont {Jordan}}, \bibinfo {author}
  {\bibfnamefont {K.~W.}\ \bibnamefont {Murch}}, \ and\ \bibinfo {author}
  {\bibfnamefont {I.}~\bibnamefont {Siddiqi}},\ }\href@noop {} {\bibfield
  {journal} {\bibinfo  {journal} {Nature}\ }\textbf {\bibinfo {volume} {511}},\
  \bibinfo {pages} {570} (\bibinfo {year} {2014})}\BibitemShut {NoStop}%
\bibitem [{\citenamefont {Jordan}\ \emph {et~al.}(2016)\citenamefont {Jordan},
  \citenamefont {Chantasri}, \citenamefont {Rouchon},\ and\ \citenamefont
  {Huard}}]{jord16}%
  \BibitemOpen
  \bibfield  {author} {\bibinfo {author} {\bibfnamefont {A.~N.}\ \bibnamefont
  {Jordan}}, \bibinfo {author} {\bibfnamefont {A.}~\bibnamefont {Chantasri}},
  \bibinfo {author} {\bibfnamefont {P.}~\bibnamefont {Rouchon}}, \ and\
  \bibinfo {author} {\bibfnamefont {B.}~\bibnamefont {Huard}},\ }\href@noop {}
  {\bibfield  {journal} {\bibinfo  {journal} {Quantum Studies: Mathematics and
  Foundations}\ }\textbf {\bibinfo {volume} {3}},\ \bibinfo {pages} {237}
  (\bibinfo {year} {2016})}\BibitemShut {NoStop}%
\bibitem [{\citenamefont {{Chantasri}}\ \emph {et~al.}(2018)\citenamefont
  {{Chantasri}}, \citenamefont {{Atalaya}}, \citenamefont {{Hacohen-Gourgy}},
  \citenamefont {{Martin}}, \citenamefont {{Siddiqi}},\ and\ \citenamefont
  {{Jordan}}}]{aree17}%
  \BibitemOpen
  \bibfield  {author} {\bibinfo {author} {\bibfnamefont {A.}~\bibnamefont
  {{Chantasri}}}, \bibinfo {author} {\bibfnamefont {J.}~\bibnamefont
  {{Atalaya}}}, \bibinfo {author} {\bibfnamefont {S.}~\bibnamefont
  {{Hacohen-Gourgy}}}, \bibinfo {author} {\bibfnamefont {L.~S.}\ \bibnamefont
  {{Martin}}}, \bibinfo {author} {\bibfnamefont {I.}~\bibnamefont {{Siddiqi}}},
  \ and\ \bibinfo {author} {\bibfnamefont {A.~N.}\ \bibnamefont {{Jordan}}},\
  }\href {\doibase 10.1103/PhysRevA.97.012118} {\bibfield  {journal} {\bibinfo
  {journal} {\pra}\ }\textbf {\bibinfo {volume} {97}},\ \bibinfo {eid} {012118}
  (\bibinfo {year} {2018})}\BibitemShut {NoStop}%
\bibitem [{\citenamefont {{Naghiloo}}\ \emph
  {et~al.}(2017{\natexlab{a}})\citenamefont {{Naghiloo}}, \citenamefont
  {{Tan}}, \citenamefont {{Harrington}}, \citenamefont {{Lewalle}},
  \citenamefont {{Jordan}},\ and\ \citenamefont {{Murch}}}]{nagh17caustic}%
  \BibitemOpen
  \bibfield  {author} {\bibinfo {author} {\bibfnamefont {M.}~\bibnamefont
  {{Naghiloo}}}, \bibinfo {author} {\bibfnamefont {D.}~\bibnamefont {{Tan}}},
  \bibinfo {author} {\bibfnamefont {P.~M.}\ \bibnamefont {{Harrington}}},
  \bibinfo {author} {\bibfnamefont {P.}~\bibnamefont {{Lewalle}}}, \bibinfo
  {author} {\bibfnamefont {A.~N.}\ \bibnamefont {{Jordan}}}, \ and\ \bibinfo
  {author} {\bibfnamefont {K.~W.}\ \bibnamefont {{Murch}}},\ }\href {\doibase
  10.1103/PhysRevA.96.053807} {\bibfield  {journal} {\bibinfo  {journal}
  {\pra}\ }\textbf {\bibinfo {volume} {96}},\ \bibinfo {eid} {053807} (\bibinfo
  {year} {2017}{\natexlab{a}})},\ \Eprint {http://arxiv.org/abs/1612.03189}
  {arXiv:1612.03189 [quant-ph]} \BibitemShut {NoStop}%
\bibitem [{\citenamefont {Crooks}(1999)}]{croo99}%
  \BibitemOpen
  \bibfield  {author} {\bibinfo {author} {\bibfnamefont {G.~E.}\ \bibnamefont
  {Crooks}},\ }\href {\doibase 10.1103/PhysRevE.60.2721} {\bibfield  {journal}
  {\bibinfo  {journal} {Phys. Rev. E}\ }\textbf {\bibinfo {volume} {60}},\
  \bibinfo {pages} {2721} (\bibinfo {year} {1999})}\BibitemShut {NoStop}%
\bibitem [{\citenamefont {Crooks}(2000)}]{croo00}%
  \BibitemOpen
  \bibfield  {author} {\bibinfo {author} {\bibfnamefont {G.~E.}\ \bibnamefont
  {Crooks}},\ }\href {\doibase 10.1103/PhysRevE.61.2361} {\bibfield  {journal}
  {\bibinfo  {journal} {Phys. Rev. E}\ }\textbf {\bibinfo {volume} {61}},\
  \bibinfo {pages} {2361} (\bibinfo {year} {2000})}\BibitemShut {NoStop}%
\bibitem [{\citenamefont {Seifert}(2005)}]{seif05}%
  \BibitemOpen
  \bibfield  {author} {\bibinfo {author} {\bibfnamefont {U.}~\bibnamefont
  {Seifert}},\ }\href {\doibase 10.1103/PhysRevLett.95.040602} {\bibfield
  {journal} {\bibinfo  {journal} {Phys. Rev. Lett.}\ }\textbf {\bibinfo
  {volume} {95}},\ \bibinfo {pages} {040602} (\bibinfo {year}
  {2005})}\BibitemShut {NoStop}%
\bibitem [{\citenamefont {{Harris}}\ and\ \citenamefont
  {{Sch{\"u}tz}}(2007)}]{harr07}%
  \BibitemOpen
  \bibfield  {author} {\bibinfo {author} {\bibfnamefont {R.~J.}\ \bibnamefont
  {{Harris}}}\ and\ \bibinfo {author} {\bibfnamefont {G.~M.}\ \bibnamefont
  {{Sch{\"u}tz}}},\ }\href {\doibase 10.1088/1742-5468/2007/07/P07020}
  {\bibfield  {journal} {\bibinfo  {journal} {Journal of Statistical Mechanics:
  Theory and Experiment}\ }\textbf {\bibinfo {volume} {7}},\ \bibinfo {pages}
  {07020} (\bibinfo {year} {2007})}\BibitemShut {NoStop}%
\bibitem [{\citenamefont {Horowitz}(2012)}]{horo11}%
  \BibitemOpen
  \bibfield  {author} {\bibinfo {author} {\bibfnamefont {J.~M.}\ \bibnamefont
  {Horowitz}},\ }\href {\doibase 10.1103/PhysRevE.85.031110} {\bibfield
  {journal} {\bibinfo  {journal} {Phys. Rev. E}\ }\textbf {\bibinfo {volume}
  {85}},\ \bibinfo {pages} {031110} (\bibinfo {year} {2012})}\BibitemShut
  {NoStop}%
\bibitem [{\citenamefont {Liphardt}\ \emph {et~al.}(2002)\citenamefont
  {Liphardt}, \citenamefont {Dumont}, \citenamefont {Smith}, \citenamefont
  {Tinoco},\ and\ \citenamefont {Bustamante}}]{lip02}%
  \BibitemOpen
  \bibfield  {author} {\bibinfo {author} {\bibfnamefont {J.}~\bibnamefont
  {Liphardt}}, \bibinfo {author} {\bibfnamefont {S.}~\bibnamefont {Dumont}},
  \bibinfo {author} {\bibfnamefont {S.~B.}\ \bibnamefont {Smith}}, \bibinfo
  {author} {\bibfnamefont {I.}~\bibnamefont {Tinoco}}, \ and\ \bibinfo {author}
  {\bibfnamefont {C.}~\bibnamefont {Bustamante}},\ }\href@noop {} {\bibfield
  {journal} {\bibinfo  {journal} {Science}\ }\textbf {\bibinfo {volume}
  {296}},\ \bibinfo {pages} {1832} (\bibinfo {year} {2002})}\BibitemShut
  {NoStop}%
\bibitem [{\citenamefont {Wang}\ \emph {et~al.}(2002)\citenamefont {Wang},
  \citenamefont {Sevick}, \citenamefont {Mittag}, \citenamefont {Searles},\
  and\ \citenamefont {Evans}}]{wan02}%
  \BibitemOpen
  \bibfield  {author} {\bibinfo {author} {\bibfnamefont {G.~M.}\ \bibnamefont
  {Wang}}, \bibinfo {author} {\bibfnamefont {E.~M.}\ \bibnamefont {Sevick}},
  \bibinfo {author} {\bibfnamefont {E.}~\bibnamefont {Mittag}}, \bibinfo
  {author} {\bibfnamefont {D.~J.}\ \bibnamefont {Searles}}, \ and\ \bibinfo
  {author} {\bibfnamefont {D.~J.}\ \bibnamefont {Evans}},\ }\href@noop {}
  {\bibfield  {journal} {\bibinfo  {journal} {Phys. Rev. Lett.}\ }\textbf
  {\bibinfo {volume} {89}},\ \bibinfo {pages} {050601} (\bibinfo {year}
  {2002})}\BibitemShut {NoStop}%
\bibitem [{\citenamefont {Tietz}\ \emph {et~al.}(2006)\citenamefont {Tietz},
  \citenamefont {Schuler}, \citenamefont {Speck}, \citenamefont {Seifert},\
  and\ \citenamefont {Wrachtrup}}]{tiet06}%
  \BibitemOpen
  \bibfield  {author} {\bibinfo {author} {\bibfnamefont {C.}~\bibnamefont
  {Tietz}}, \bibinfo {author} {\bibfnamefont {S.}~\bibnamefont {Schuler}},
  \bibinfo {author} {\bibfnamefont {T.}~\bibnamefont {Speck}}, \bibinfo
  {author} {\bibfnamefont {U.}~\bibnamefont {Seifert}}, \ and\ \bibinfo
  {author} {\bibfnamefont {J.}~\bibnamefont {Wrachtrup}},\ }\href {\doibase
  10.1103/PhysRevLett.97.050602} {\bibfield  {journal} {\bibinfo  {journal}
  {Phys. Rev. Lett.}\ }\textbf {\bibinfo {volume} {97}},\ \bibinfo {pages}
  {050602} (\bibinfo {year} {2006})}\BibitemShut {NoStop}%
\bibitem [{\citenamefont {Esposito}\ \emph {et~al.}(2007)\citenamefont
  {Esposito}, \citenamefont {Harbola},\ and\ \citenamefont {Mukamel}}]{espo07}%
  \BibitemOpen
  \bibfield  {author} {\bibinfo {author} {\bibfnamefont {M.}~\bibnamefont
  {Esposito}}, \bibinfo {author} {\bibfnamefont {U.}~\bibnamefont {Harbola}}, \
  and\ \bibinfo {author} {\bibfnamefont {S.}~\bibnamefont {Mukamel}},\ }\href
  {\doibase 10.1103/PhysRevE.76.031132} {\bibfield  {journal} {\bibinfo
  {journal} {Phys. Rev. E}\ }\textbf {\bibinfo {volume} {76}},\ \bibinfo
  {pages} {031132} (\bibinfo {year} {2007})}\BibitemShut {NoStop}%
\bibitem [{\citenamefont {Blickle}\ \emph {et~al.}(2006)\citenamefont
  {Blickle}, \citenamefont {Speck}, \citenamefont {Helden}, \citenamefont
  {Seifert},\ and\ \citenamefont {Bechinger}}]{bli06}%
  \BibitemOpen
  \bibfield  {author} {\bibinfo {author} {\bibfnamefont {V.}~\bibnamefont
  {Blickle}}, \bibinfo {author} {\bibfnamefont {T.}~\bibnamefont {Speck}},
  \bibinfo {author} {\bibfnamefont {L.}~\bibnamefont {Helden}}, \bibinfo
  {author} {\bibfnamefont {U.}~\bibnamefont {Seifert}}, \ and\ \bibinfo
  {author} {\bibfnamefont {C.}~\bibnamefont {Bechinger}},\ }\href@noop {}
  {\bibfield  {journal} {\bibinfo  {journal} {Phys. Rev. Lett.}\ }\textbf
  {\bibinfo {volume} {96}},\ \bibinfo {pages} {070603} (\bibinfo {year}
  {2006})}\BibitemShut {NoStop}%
\bibitem [{\citenamefont {Speck}\ \emph {et~al.}(2007)\citenamefont {Speck},
  \citenamefont {Blickle}, \citenamefont {Bechinger},\ and\ \citenamefont
  {Seifert}}]{spec07}%
  \BibitemOpen
  \bibfield  {author} {\bibinfo {author} {\bibfnamefont {T.}~\bibnamefont
  {Speck}}, \bibinfo {author} {\bibfnamefont {V.}~\bibnamefont {Blickle}},
  \bibinfo {author} {\bibfnamefont {C.}~\bibnamefont {Bechinger}}, \ and\
  \bibinfo {author} {\bibfnamefont {U.}~\bibnamefont {Seifert}},\ }\href@noop
  {} {\bibfield  {journal} {\bibinfo  {journal} {EPL (Europhysics Letters)}\
  }\textbf {\bibinfo {volume} {79}},\ \bibinfo {pages} {30002} (\bibinfo {year}
  {2007})}\BibitemShut {NoStop}%
\bibitem [{\citenamefont {Utsumi}\ \emph {et~al.}(2010)\citenamefont {Utsumi},
  \citenamefont {Golubev}, \citenamefont {Marthaler}, \citenamefont {Saito},
  \citenamefont {Fujisawa},\ and\ \citenamefont {Sch\"on}}]{utsu10}%
  \BibitemOpen
  \bibfield  {author} {\bibinfo {author} {\bibfnamefont {Y.}~\bibnamefont
  {Utsumi}}, \bibinfo {author} {\bibfnamefont {D.~S.}\ \bibnamefont {Golubev}},
  \bibinfo {author} {\bibfnamefont {M.}~\bibnamefont {Marthaler}}, \bibinfo
  {author} {\bibfnamefont {K.}~\bibnamefont {Saito}}, \bibinfo {author}
  {\bibfnamefont {T.}~\bibnamefont {Fujisawa}}, \ and\ \bibinfo {author}
  {\bibfnamefont {G.}~\bibnamefont {Sch\"on}},\ }\href {\doibase
  10.1103/PhysRevB.81.125331} {\bibfield  {journal} {\bibinfo  {journal} {Phys.
  Rev. B}\ }\textbf {\bibinfo {volume} {81}},\ \bibinfo {pages} {125331}
  (\bibinfo {year} {2010})}\BibitemShut {NoStop}%
\bibitem [{\citenamefont {K\"ung}\ \emph {et~al.}(2012)\citenamefont {K\"ung},
  \citenamefont {R\"ossler}, \citenamefont {Beck}, \citenamefont {Marthaler},
  \citenamefont {Golubev}, \citenamefont {Utsumi}, \citenamefont {Ihn},\ and\
  \citenamefont {Ensslin}}]{kung12}%
  \BibitemOpen
  \bibfield  {author} {\bibinfo {author} {\bibfnamefont {B.}~\bibnamefont
  {K\"ung}}, \bibinfo {author} {\bibfnamefont {C.}~\bibnamefont {R\"ossler}},
  \bibinfo {author} {\bibfnamefont {M.}~\bibnamefont {Beck}}, \bibinfo {author}
  {\bibfnamefont {M.}~\bibnamefont {Marthaler}}, \bibinfo {author}
  {\bibfnamefont {D.~S.}\ \bibnamefont {Golubev}}, \bibinfo {author}
  {\bibfnamefont {Y.}~\bibnamefont {Utsumi}}, \bibinfo {author} {\bibfnamefont
  {T.}~\bibnamefont {Ihn}}, \ and\ \bibinfo {author} {\bibfnamefont
  {K.}~\bibnamefont {Ensslin}},\ }\href {\doibase 10.1103/PhysRevX.2.011001}
  {\bibfield  {journal} {\bibinfo  {journal} {Phys. Rev. X}\ }\textbf {\bibinfo
  {volume} {2}},\ \bibinfo {pages} {011001} (\bibinfo {year}
  {2012})}\BibitemShut {NoStop}%
\bibitem [{\citenamefont {Saira}\ \emph {et~al.}(2012)\citenamefont {Saira},
  \citenamefont {Yoon}, \citenamefont {Tanttu}, \citenamefont {M\"ott\"onen},
  \citenamefont {Averin},\ and\ \citenamefont {Pekola}}]{sair12}%
  \BibitemOpen
  \bibfield  {author} {\bibinfo {author} {\bibfnamefont {O.-P.}\ \bibnamefont
  {Saira}}, \bibinfo {author} {\bibfnamefont {Y.}~\bibnamefont {Yoon}},
  \bibinfo {author} {\bibfnamefont {T.}~\bibnamefont {Tanttu}}, \bibinfo
  {author} {\bibfnamefont {M.}~\bibnamefont {M\"ott\"onen}}, \bibinfo {author}
  {\bibfnamefont {D.~V.}\ \bibnamefont {Averin}}, \ and\ \bibinfo {author}
  {\bibfnamefont {J.~P.}\ \bibnamefont {Pekola}},\ }\href {\doibase
  10.1103/PhysRevLett.109.180601} {\bibfield  {journal} {\bibinfo  {journal}
  {Phys. Rev. Lett.}\ }\textbf {\bibinfo {volume} {109}},\ \bibinfo {pages}
  {180601} (\bibinfo {year} {2012})}\BibitemShut {NoStop}%
\bibitem [{\citenamefont {Koski}\ \emph {et~al.}(2013)\citenamefont {Koski},
  \citenamefont {Sagawa}, \citenamefont {Saira}, \citenamefont {Yoon},
  \citenamefont {Kutvonen}, \citenamefont {Solinas}, \citenamefont
  {M\"ott\"onen}, \citenamefont {Ala-Nissila},\ and\ \citenamefont
  {Pekola}}]{kosk13}%
  \BibitemOpen
  \bibfield  {author} {\bibinfo {author} {\bibfnamefont {J.~V.}\ \bibnamefont
  {Koski}}, \bibinfo {author} {\bibfnamefont {T.}~\bibnamefont {Sagawa}},
  \bibinfo {author} {\bibfnamefont {O.-P.}\ \bibnamefont {Saira}}, \bibinfo
  {author} {\bibfnamefont {Y.}~\bibnamefont {Yoon}}, \bibinfo {author}
  {\bibfnamefont {A.}~\bibnamefont {Kutvonen}}, \bibinfo {author}
  {\bibfnamefont {P.}~\bibnamefont {Solinas}}, \bibinfo {author} {\bibfnamefont
  {M.}~\bibnamefont {M\"ott\"onen}}, \bibinfo {author} {\bibfnamefont
  {T.}~\bibnamefont {Ala-Nissila}}, \ and\ \bibinfo {author} {\bibfnamefont
  {J.~P.}\ \bibnamefont {Pekola}},\ }\href@noop {} {\bibfield  {journal}
  {\bibinfo  {journal} {Nature Physics}\ }\textbf {\bibinfo {volume} {9}},\
  \bibinfo {pages} {644 EP } (\bibinfo {year} {2013})}\BibitemShut {NoStop}%
\bibitem [{\citenamefont {Hoang}\ \emph {et~al.}(2018)\citenamefont {Hoang},
  \citenamefont {Pan}, \citenamefont {Ahn}, \citenamefont {Bang}, \citenamefont
  {Quan},\ and\ \citenamefont {Li}}]{hoan18}%
  \BibitemOpen
  \bibfield  {author} {\bibinfo {author} {\bibfnamefont {T.~M.}\ \bibnamefont
  {Hoang}}, \bibinfo {author} {\bibfnamefont {R.}~\bibnamefont {Pan}}, \bibinfo
  {author} {\bibfnamefont {J.}~\bibnamefont {Ahn}}, \bibinfo {author}
  {\bibfnamefont {J.}~\bibnamefont {Bang}}, \bibinfo {author} {\bibfnamefont
  {H.~T.}\ \bibnamefont {Quan}}, \ and\ \bibinfo {author} {\bibfnamefont
  {T.}~\bibnamefont {Li}},\ }\href {\doibase 10.1103/PhysRevLett.120.080602}
  {\bibfield  {journal} {\bibinfo  {journal} {Phys. Rev. Lett.}\ }\textbf
  {\bibinfo {volume} {120}},\ \bibinfo {pages} {080602} (\bibinfo {year}
  {2018})}\BibitemShut {NoStop}%
\bibitem [{\citenamefont {Jarzynski}(1997)}]{jarz97}%
  \BibitemOpen
  \bibfield  {author} {\bibinfo {author} {\bibfnamefont {C.}~\bibnamefont
  {Jarzynski}},\ }\href {\doibase 10.1103/PhysRevLett.78.2690} {\bibfield
  {journal} {\bibinfo  {journal} {Phys. Rev. Lett.}\ }\textbf {\bibinfo
  {volume} {78}},\ \bibinfo {pages} {2690} (\bibinfo {year}
  {1997})}\BibitemShut {NoStop}%
\bibitem [{\citenamefont {Crooks}(1998)}]{croo98}%
  \BibitemOpen
  \bibfield  {author} {\bibinfo {author} {\bibfnamefont {G.~E.}\ \bibnamefont
  {Crooks}},\ }\href@noop {} {\bibfield  {journal} {\bibinfo  {journal}
  {Journal of Statistical Physics}\ }\textbf {\bibinfo {volume} {90}},\
  \bibinfo {pages} {1481} (\bibinfo {year} {1998})}\BibitemShut {NoStop}%
\bibitem [{\citenamefont {Esposito}\ and\ \citenamefont {Van~den
  Broeck}(2010)}]{espo10}%
  \BibitemOpen
  \bibfield  {author} {\bibinfo {author} {\bibfnamefont {M.}~\bibnamefont
  {Esposito}}\ and\ \bibinfo {author} {\bibfnamefont {C.}~\bibnamefont {Van~den
  Broeck}},\ }\href {\doibase 10.1103/PhysRevLett.104.090601} {\bibfield
  {journal} {\bibinfo  {journal} {Phys. Rev. Lett.}\ }\textbf {\bibinfo
  {volume} {104}},\ \bibinfo {pages} {090601} (\bibinfo {year}
  {2010})}\BibitemShut {NoStop}%
\bibitem [{\citenamefont {Jarzynski}(2011)}]{jar11}%
  \BibitemOpen
  \bibfield  {author} {\bibinfo {author} {\bibfnamefont {C.}~\bibnamefont
  {Jarzynski}},\ }\href@noop {} {\bibfield  {journal} {\bibinfo  {journal} {C.
  Annu. Rev. Condens. Matter Phys.}\ }\textbf {\bibinfo {volume} {2}},\
  \bibinfo {pages} {329} (\bibinfo {year} {2011})}\BibitemShut {NoStop}%
\bibitem [{\citenamefont {Seifert}(2012)}]{seif12}%
  \BibitemOpen
  \bibfield  {author} {\bibinfo {author} {\bibfnamefont {U.}~\bibnamefont
  {Seifert}},\ }\href@noop {} {\bibfield  {journal} {\bibinfo  {journal} {Rep.
  Prog. Phys.}\ }\textbf {\bibinfo {volume} {75}},\ \bibinfo {pages} {126001}
  (\bibinfo {year} {2012})}\BibitemShut {NoStop}%
\bibitem [{\citenamefont {Ciliberto}\ \emph {et~al.}(2013)\citenamefont
  {Ciliberto}, \citenamefont {Gomez-Solano},\ and\ \citenamefont
  {Petrosyan}}]{cil13}%
  \BibitemOpen
  \bibfield  {author} {\bibinfo {author} {\bibfnamefont {S.}~\bibnamefont
  {Ciliberto}}, \bibinfo {author} {\bibfnamefont {R.}~\bibnamefont
  {Gomez-Solano}}, \ and\ \bibinfo {author} {\bibfnamefont {A.}~\bibnamefont
  {Petrosyan}},\ }\href@noop {} {\bibfield  {journal} {\bibinfo  {journal} {C.
  Annu. Rev. Condens. Matter Phys.}\ }\textbf {\bibinfo {volume} {4}},\
  \bibinfo {pages} {235} (\bibinfo {year} {2013})}\BibitemShut {NoStop}%
\bibitem [{\citenamefont {Manzano}\ \emph {et~al.}(2018)\citenamefont
  {Manzano}, \citenamefont {Horowitz},\ and\ \citenamefont
  {Parrondo}}]{manz18}%
  \BibitemOpen
  \bibfield  {author} {\bibinfo {author} {\bibfnamefont {G.}~\bibnamefont
  {Manzano}}, \bibinfo {author} {\bibfnamefont {J.~M.}\ \bibnamefont
  {Horowitz}}, \ and\ \bibinfo {author} {\bibfnamefont {J.~M.~R.}\ \bibnamefont
  {Parrondo}},\ }\href {\doibase 10.1103/PhysRevX.8.031037} {\bibfield
  {journal} {\bibinfo  {journal} {Phys. Rev. X}\ }\textbf {\bibinfo {volume}
  {8}},\ \bibinfo {pages} {031037} (\bibinfo {year} {2018})}\BibitemShut
  {NoStop}%
\bibitem [{\citenamefont {{Kurchan}}(2000)}]{kurc01}%
  \BibitemOpen
  \bibfield  {author} {\bibinfo {author} {\bibfnamefont {J.}~\bibnamefont
  {{Kurchan}}},\ }\href@noop {} {\bibfield  {journal} {\bibinfo  {journal}
  {arXiv:cond-mat/0007360}\ } (\bibinfo {year} {2000})}\BibitemShut {NoStop}%
\bibitem [{\citenamefont {Mukamel}(2003)}]{muka03}%
  \BibitemOpen
  \bibfield  {author} {\bibinfo {author} {\bibfnamefont {S.}~\bibnamefont
  {Mukamel}},\ }\href {\doibase 10.1103/PhysRevLett.90.170604} {\bibfield
  {journal} {\bibinfo  {journal} {Phys. Rev. Lett.}\ }\textbf {\bibinfo
  {volume} {90}},\ \bibinfo {pages} {170604} (\bibinfo {year}
  {2003})}\BibitemShut {NoStop}%
\bibitem [{\citenamefont {Campisi}\ \emph {et~al.}(2009)\citenamefont
  {Campisi}, \citenamefont {Talkner},\ and\ \citenamefont {H\"anggi}}]{camp09}%
  \BibitemOpen
  \bibfield  {author} {\bibinfo {author} {\bibfnamefont {M.}~\bibnamefont
  {Campisi}}, \bibinfo {author} {\bibfnamefont {P.}~\bibnamefont {Talkner}}, \
  and\ \bibinfo {author} {\bibfnamefont {P.}~\bibnamefont {H\"anggi}},\ }\href
  {\doibase 10.1103/PhysRevLett.102.210401} {\bibfield  {journal} {\bibinfo
  {journal} {Phys. Rev. Lett.}\ }\textbf {\bibinfo {volume} {102}},\ \bibinfo
  {pages} {210401} (\bibinfo {year} {2009})}\BibitemShut {NoStop}%
\bibitem [{\citenamefont {Deffner}\ and\ \citenamefont {Lutz}(2011)}]{deff11}%
  \BibitemOpen
  \bibfield  {author} {\bibinfo {author} {\bibfnamefont {S.}~\bibnamefont
  {Deffner}}\ and\ \bibinfo {author} {\bibfnamefont {E.}~\bibnamefont {Lutz}},\
  }\href {\doibase 10.1103/PhysRevLett.107.140404} {\bibfield  {journal}
  {\bibinfo  {journal} {Phys. Rev. Lett.}\ }\textbf {\bibinfo {volume} {107}},\
  \bibinfo {pages} {140404} (\bibinfo {year} {2011})}\BibitemShut {NoStop}%
\bibitem [{\citenamefont {Morikuni}\ and\ \citenamefont
  {Tasaki}(2011)}]{mori11}%
  \BibitemOpen
  \bibfield  {author} {\bibinfo {author} {\bibfnamefont {Y.}~\bibnamefont
  {Morikuni}}\ and\ \bibinfo {author} {\bibfnamefont {H.}~\bibnamefont
  {Tasaki}},\ }\href {\doibase 10.1007/s10955-011-0153-7} {\bibfield  {journal}
  {\bibinfo  {journal} {Journal of Statistical Physics}\ }\textbf {\bibinfo
  {volume} {143}},\ \bibinfo {pages} {1} (\bibinfo {year} {2011})}\BibitemShut
  {NoStop}%
\bibitem [{\citenamefont {Chetrite}\ and\ \citenamefont
  {Mallick}(2012)}]{chet12}%
  \BibitemOpen
  \bibfield  {author} {\bibinfo {author} {\bibfnamefont {R.}~\bibnamefont
  {Chetrite}}\ and\ \bibinfo {author} {\bibfnamefont {K.}~\bibnamefont
  {Mallick}},\ }\href {\doibase 10.1007/s10955-012-0557-z} {\bibfield
  {journal} {\bibinfo  {journal} {Journal of Statistical Physics}\ }\textbf
  {\bibinfo {volume} {148}},\ \bibinfo {pages} {480} (\bibinfo {year}
  {2012})}\BibitemShut {NoStop}%
\bibitem [{\citenamefont {Horowitz}\ and\ \citenamefont
  {Parrondo}(2013)}]{horo13}%
  \BibitemOpen
  \bibfield  {author} {\bibinfo {author} {\bibfnamefont {J.~M.}\ \bibnamefont
  {Horowitz}}\ and\ \bibinfo {author} {\bibfnamefont {J.~M.~R.}\ \bibnamefont
  {Parrondo}},\ }\href {\doibase 10.1088/1367-2630/15/8/085028} {\bibfield
  {journal} {\bibinfo  {journal} {New Journal of Physics}\ }\textbf {\bibinfo
  {volume} {15}},\ \bibinfo {pages} {085028} (\bibinfo {year}
  {2013})}\BibitemShut {NoStop}%
\bibitem [{\citenamefont {Quan}\ \emph {et~al.}(2006)\citenamefont {Quan},
  \citenamefont {Wang}, \citenamefont {Liu}, \citenamefont {Sun},\ and\
  \citenamefont {Nori}}]{quan06}%
  \BibitemOpen
  \bibfield  {author} {\bibinfo {author} {\bibfnamefont {H.~T.}\ \bibnamefont
  {Quan}}, \bibinfo {author} {\bibfnamefont {Y.~D.}\ \bibnamefont {Wang}},
  \bibinfo {author} {\bibfnamefont {Y.-x.}\ \bibnamefont {Liu}}, \bibinfo
  {author} {\bibfnamefont {C.~P.}\ \bibnamefont {Sun}}, \ and\ \bibinfo
  {author} {\bibfnamefont {F.}~\bibnamefont {Nori}},\ }\href {\doibase
  10.1103/PhysRevLett.97.180402} {\bibfield  {journal} {\bibinfo  {journal}
  {Phys. Rev. Lett.}\ }\textbf {\bibinfo {volume} {97}},\ \bibinfo {pages}
  {180402} (\bibinfo {year} {2006})}\BibitemShut {NoStop}%
\bibitem [{\citenamefont {Quan}\ \emph {et~al.}(2007)\citenamefont {Quan},
  \citenamefont {Liu}, \citenamefont {Sun},\ and\ \citenamefont
  {Nori}}]{quan07}%
  \BibitemOpen
  \bibfield  {author} {\bibinfo {author} {\bibfnamefont {H.~T.}\ \bibnamefont
  {Quan}}, \bibinfo {author} {\bibfnamefont {Y.-x.}\ \bibnamefont {Liu}},
  \bibinfo {author} {\bibfnamefont {C.~P.}\ \bibnamefont {Sun}}, \ and\
  \bibinfo {author} {\bibfnamefont {F.}~\bibnamefont {Nori}},\ }\href {\doibase
  10.1103/PhysRevE.76.031105} {\bibfield  {journal} {\bibinfo  {journal} {Phys.
  Rev. E}\ }\textbf {\bibinfo {volume} {76}},\ \bibinfo {pages} {031105}
  (\bibinfo {year} {2007})}\BibitemShut {NoStop}%
\bibitem [{\citenamefont {Dorner}\ \emph {et~al.}(2013)\citenamefont {Dorner},
  \citenamefont {Clark}, \citenamefont {Heaney}, \citenamefont {Fazio},
  \citenamefont {Goold},\ and\ \citenamefont {Vedral}}]{dorn13}%
  \BibitemOpen
  \bibfield  {author} {\bibinfo {author} {\bibfnamefont {R.}~\bibnamefont
  {Dorner}}, \bibinfo {author} {\bibfnamefont {S.~R.}\ \bibnamefont {Clark}},
  \bibinfo {author} {\bibfnamefont {L.}~\bibnamefont {Heaney}}, \bibinfo
  {author} {\bibfnamefont {R.}~\bibnamefont {Fazio}}, \bibinfo {author}
  {\bibfnamefont {J.}~\bibnamefont {Goold}}, \ and\ \bibinfo {author}
  {\bibfnamefont {V.}~\bibnamefont {Vedral}},\ }\href {\doibase
  10.1103/PhysRevLett.110.230601} {\bibfield  {journal} {\bibinfo  {journal}
  {Phys. Rev. Lett.}\ }\textbf {\bibinfo {volume} {110}},\ \bibinfo {pages}
  {230601} (\bibinfo {year} {2013})}\BibitemShut {NoStop}%
\bibitem [{\citenamefont {Mazzola}\ \emph {et~al.}(2013)\citenamefont
  {Mazzola}, \citenamefont {De~Chiara},\ and\ \citenamefont
  {Paternostro}}]{mazz13}%
  \BibitemOpen
  \bibfield  {author} {\bibinfo {author} {\bibfnamefont {L.}~\bibnamefont
  {Mazzola}}, \bibinfo {author} {\bibfnamefont {G.}~\bibnamefont {De~Chiara}},
  \ and\ \bibinfo {author} {\bibfnamefont {M.}~\bibnamefont {Paternostro}},\
  }\href {\doibase 10.1103/PhysRevLett.110.230602} {\bibfield  {journal}
  {\bibinfo  {journal} {Phys. Rev. Lett.}\ }\textbf {\bibinfo {volume} {110}},\
  \bibinfo {pages} {230602} (\bibinfo {year} {2013})}\BibitemShut {NoStop}%
\bibitem [{\citenamefont {Campisi}\ \emph {et~al.}(2013)\citenamefont
  {Campisi}, \citenamefont {Blattmann}, \citenamefont {Kohler}, \citenamefont
  {Zueco},\ and\ \citenamefont {H\"anggi}}]{camp13}%
  \BibitemOpen
  \bibfield  {author} {\bibinfo {author} {\bibfnamefont {M.}~\bibnamefont
  {Campisi}}, \bibinfo {author} {\bibfnamefont {R.}~\bibnamefont {Blattmann}},
  \bibinfo {author} {\bibfnamefont {S.}~\bibnamefont {Kohler}}, \bibinfo
  {author} {\bibfnamefont {D.}~\bibnamefont {Zueco}}, \ and\ \bibinfo {author}
  {\bibfnamefont {P.}~\bibnamefont {H\"anggi}},\ }\href {\doibase
  10.1088/1367-2630/15/10/105028} {\bibfield  {journal} {\bibinfo  {journal}
  {New Journal of Physics}\ }\textbf {\bibinfo {volume} {15}},\ \bibinfo
  {pages} {105028} (\bibinfo {year} {2013})}\BibitemShut {NoStop}%
\bibitem [{\citenamefont {Goold}\ \emph {et~al.}(2014)\citenamefont {Goold},
  \citenamefont {Poschinger},\ and\ \citenamefont {Modi}}]{gool14}%
  \BibitemOpen
  \bibfield  {author} {\bibinfo {author} {\bibfnamefont {J.}~\bibnamefont
  {Goold}}, \bibinfo {author} {\bibfnamefont {U.}~\bibnamefont {Poschinger}}, \
  and\ \bibinfo {author} {\bibfnamefont {K.}~\bibnamefont {Modi}},\ }\href
  {\doibase 10.1103/PhysRevE.90.020101} {\bibfield  {journal} {\bibinfo
  {journal} {Phys. Rev. E}\ }\textbf {\bibinfo {volume} {90}},\ \bibinfo
  {pages} {020101} (\bibinfo {year} {2014})}\BibitemShut {NoStop}%
\bibitem [{\citenamefont {Roncaglia}\ \emph {et~al.}(2014)\citenamefont
  {Roncaglia}, \citenamefont {Cerisola},\ and\ \citenamefont {Paz}}]{ronc14}%
  \BibitemOpen
  \bibfield  {author} {\bibinfo {author} {\bibfnamefont {A.~J.}\ \bibnamefont
  {Roncaglia}}, \bibinfo {author} {\bibfnamefont {F.}~\bibnamefont {Cerisola}},
  \ and\ \bibinfo {author} {\bibfnamefont {J.~P.}\ \bibnamefont {Paz}},\ }\href
  {\doibase 10.1103/PhysRevLett.113.250601} {\bibfield  {journal} {\bibinfo
  {journal} {Phys. Rev. Lett.}\ }\textbf {\bibinfo {volume} {113}},\ \bibinfo
  {pages} {250601} (\bibinfo {year} {2014})}\BibitemShut {NoStop}%
\bibitem [{\citenamefont {Chiara}\ \emph {et~al.}(2015)\citenamefont {Chiara},
  \citenamefont {Roncaglia},\ and\ \citenamefont {Paz}}]{chia15}%
  \BibitemOpen
  \bibfield  {author} {\bibinfo {author} {\bibfnamefont {G.~D.}\ \bibnamefont
  {Chiara}}, \bibinfo {author} {\bibfnamefont {A.~J.}\ \bibnamefont
  {Roncaglia}}, \ and\ \bibinfo {author} {\bibfnamefont {J.~P.}\ \bibnamefont
  {Paz}},\ }\href {\doibase 10.1088/1367-2630/17/3/035004} {\bibfield
  {journal} {\bibinfo  {journal} {New Journal of Physics}\ }\textbf {\bibinfo
  {volume} {17}},\ \bibinfo {pages} {035004} (\bibinfo {year}
  {2015})}\BibitemShut {NoStop}%
\bibitem [{\citenamefont {Batalh\~ao}\ \emph {et~al.}(2015)\citenamefont
  {Batalh\~ao}, \citenamefont {Souza}, \citenamefont {Sarthour}, \citenamefont
  {Oliveira}, \citenamefont {Paternostro}, \citenamefont {Lutz},\ and\
  \citenamefont {Serra}}]{bata15}%
  \BibitemOpen
  \bibfield  {author} {\bibinfo {author} {\bibfnamefont {T.~B.}\ \bibnamefont
  {Batalh\~ao}}, \bibinfo {author} {\bibfnamefont {A.~M.}\ \bibnamefont
  {Souza}}, \bibinfo {author} {\bibfnamefont {R.~S.}\ \bibnamefont {Sarthour}},
  \bibinfo {author} {\bibfnamefont {I.~S.}\ \bibnamefont {Oliveira}}, \bibinfo
  {author} {\bibfnamefont {M.}~\bibnamefont {Paternostro}}, \bibinfo {author}
  {\bibfnamefont {E.}~\bibnamefont {Lutz}}, \ and\ \bibinfo {author}
  {\bibfnamefont {R.~M.}\ \bibnamefont {Serra}},\ }\href {\doibase
  10.1103/PhysRevLett.115.190601} {\bibfield  {journal} {\bibinfo  {journal}
  {Phys. Rev. Lett.}\ }\textbf {\bibinfo {volume} {115}},\ \bibinfo {pages}
  {190601} (\bibinfo {year} {2015})}\BibitemShut {NoStop}%
\bibitem [{\citenamefont {{An}}\ \emph {et~al.}(2015)\citenamefont {{An}},
  \citenamefont {{Zhang}}, \citenamefont {{Um}}, \citenamefont {{Lv}},
  \citenamefont {{Lu}}, \citenamefont {{Zhang}}, \citenamefont {{Yin}},
  \citenamefont {{Quan}},\ and\ \citenamefont {{Kim}}}]{an15}%
  \BibitemOpen
  \bibfield  {author} {\bibinfo {author} {\bibfnamefont {S.}~\bibnamefont
  {{An}}}, \bibinfo {author} {\bibfnamefont {J.-N.}\ \bibnamefont {{Zhang}}},
  \bibinfo {author} {\bibfnamefont {M.}~\bibnamefont {{Um}}}, \bibinfo {author}
  {\bibfnamefont {D.}~\bibnamefont {{Lv}}}, \bibinfo {author} {\bibfnamefont
  {Y.}~\bibnamefont {{Lu}}}, \bibinfo {author} {\bibfnamefont {J.}~\bibnamefont
  {{Zhang}}}, \bibinfo {author} {\bibfnamefont {Z.-Q.}\ \bibnamefont {{Yin}}},
  \bibinfo {author} {\bibfnamefont {H.~T.}\ \bibnamefont {{Quan}}}, \ and\
  \bibinfo {author} {\bibfnamefont {K.}~\bibnamefont {{Kim}}},\ }\href
  {\doibase 10.1038/nphys3197} {\bibfield  {journal} {\bibinfo  {journal}
  {Nature Physics}\ }\textbf {\bibinfo {volume} {11}},\ \bibinfo {pages} {193}
  (\bibinfo {year} {2015})},\ \Eprint {http://arxiv.org/abs/1409.4485}
  {arXiv:1409.4485 [quant-ph]} \BibitemShut {NoStop}%
\bibitem [{\citenamefont {Alonso}\ \emph {et~al.}(2016)\citenamefont {Alonso},
  \citenamefont {Lutz},\ and\ \citenamefont {Romito}}]{alon16}%
  \BibitemOpen
  \bibfield  {author} {\bibinfo {author} {\bibfnamefont {J.~J.}\ \bibnamefont
  {Alonso}}, \bibinfo {author} {\bibfnamefont {E.}~\bibnamefont {Lutz}}, \ and\
  \bibinfo {author} {\bibfnamefont {A.}~\bibnamefont {Romito}},\ }\href
  {\doibase 10.1103/PhysRevLett.116.080403} {\bibfield  {journal} {\bibinfo
  {journal} {Phys. Rev. Lett.}\ }\textbf {\bibinfo {volume} {116}},\ \bibinfo
  {pages} {080403} (\bibinfo {year} {2016})}\BibitemShut {NoStop}%
\bibitem [{\citenamefont {Watanabe}\ \emph
  {et~al.}(2014{\natexlab{a}})\citenamefont {Watanabe}, \citenamefont
  {Venkatesh},\ and\ \citenamefont {Talkner}}]{wata14}%
  \BibitemOpen
  \bibfield  {author} {\bibinfo {author} {\bibfnamefont {G.}~\bibnamefont
  {Watanabe}}, \bibinfo {author} {\bibfnamefont {B.~P.}\ \bibnamefont
  {Venkatesh}}, \ and\ \bibinfo {author} {\bibfnamefont {P.}~\bibnamefont
  {Talkner}},\ }\href {\doibase 10.1103/PhysRevE.89.052116} {\bibfield
  {journal} {\bibinfo  {journal} {Phys. Rev. E}\ }\textbf {\bibinfo {volume}
  {89}},\ \bibinfo {pages} {052116} (\bibinfo {year}
  {2014}{\natexlab{a}})}\BibitemShut {NoStop}%
\bibitem [{\citenamefont {Yi}\ \emph {et~al.}(2017)\citenamefont {Yi},
  \citenamefont {Talkner},\ and\ \citenamefont {Kim}}]{yi14}%
  \BibitemOpen
  \bibfield  {author} {\bibinfo {author} {\bibfnamefont {J.}~\bibnamefont
  {Yi}}, \bibinfo {author} {\bibfnamefont {P.}~\bibnamefont {Talkner}}, \ and\
  \bibinfo {author} {\bibfnamefont {Y.~W.}\ \bibnamefont {Kim}},\ }\href
  {\doibase 10.1103/PhysRevE.96.022108} {\bibfield  {journal} {\bibinfo
  {journal} {Phys. Rev. E}\ }\textbf {\bibinfo {volume} {96}},\ \bibinfo
  {pages} {022108} (\bibinfo {year} {2017})}\BibitemShut {NoStop}%
\bibitem [{\citenamefont {Campisi}\ \emph {et~al.}(2011)\citenamefont
  {Campisi}, \citenamefont {H\"anggi},\ and\ \citenamefont {Talkner}}]{camp11}%
  \BibitemOpen
  \bibfield  {author} {\bibinfo {author} {\bibfnamefont {M.}~\bibnamefont
  {Campisi}}, \bibinfo {author} {\bibfnamefont {P.}~\bibnamefont {H\"anggi}}, \
  and\ \bibinfo {author} {\bibfnamefont {P.}~\bibnamefont {Talkner}},\ }\href
  {\doibase 10.1103/RevModPhys.83.771} {\bibfield  {journal} {\bibinfo
  {journal} {Rev. Mod. Phys.}\ }\textbf {\bibinfo {volume} {83}},\ \bibinfo
  {pages} {771} (\bibinfo {year} {2011})}\BibitemShut {NoStop}%
\bibitem [{\citenamefont {Watanabe}\ \emph
  {et~al.}(2014{\natexlab{b}})\citenamefont {Watanabe}, \citenamefont
  {Venkatesh},\ and\ \citenamefont {Talkner}}]{wata14_generalized}%
  \BibitemOpen
  \bibfield  {author} {\bibinfo {author} {\bibfnamefont {G.}~\bibnamefont
  {Watanabe}}, \bibinfo {author} {\bibfnamefont {B.~P.}\ \bibnamefont
  {Venkatesh}}, \ and\ \bibinfo {author} {\bibfnamefont {P.}~\bibnamefont
  {Talkner}},\ }\href {\doibase 10.1103/PhysRevE.89.052116} {\bibfield
  {journal} {\bibinfo  {journal} {Phys. Rev. E}\ }\textbf {\bibinfo {volume}
  {89}},\ \bibinfo {pages} {052116} (\bibinfo {year}
  {2014}{\natexlab{b}})}\BibitemShut {NoStop}%
\bibitem [{\citenamefont {Elouard}\ \emph
  {et~al.}(2017{\natexlab{a}})\citenamefont {Elouard}, \citenamefont
  {Herrera-Mart{\'{\i}}}, \citenamefont {Clusel},\ and\ \citenamefont
  {Auff{\`{e}}ves}}]{elou17_role}%
  \BibitemOpen
  \bibfield  {author} {\bibinfo {author} {\bibfnamefont {C.}~\bibnamefont
  {Elouard}}, \bibinfo {author} {\bibfnamefont {D.~A.}\ \bibnamefont
  {Herrera-Mart{\'{\i}}}}, \bibinfo {author} {\bibfnamefont {M.}~\bibnamefont
  {Clusel}}, \ and\ \bibinfo {author} {\bibfnamefont {A.}~\bibnamefont
  {Auff{\`{e}}ves}},\ }\href {\doibase 10.1038/s41534-017-0008-4} {\bibfield
  {journal} {\bibinfo  {journal} {npj Quantum Information}\ }\textbf {\bibinfo
  {volume} {3}} (\bibinfo {year} {2017}{\natexlab{a}}),\
  10.1038/s41534-017-0008-4}\BibitemShut {NoStop}%
\bibitem [{\citenamefont {Benoist}\ \emph {et~al.}(2017)\citenamefont
  {Benoist}, \citenamefont {Jak{\v{s}}i{\'{c}}}, \citenamefont {Pautrat},\ and\
  \citenamefont {Pillet}}]{beno18}%
  \BibitemOpen
  \bibfield  {author} {\bibinfo {author} {\bibfnamefont {T.}~\bibnamefont
  {Benoist}}, \bibinfo {author} {\bibfnamefont {V.}~\bibnamefont
  {Jak{\v{s}}i{\'{c}}}}, \bibinfo {author} {\bibfnamefont {Y.}~\bibnamefont
  {Pautrat}}, \ and\ \bibinfo {author} {\bibfnamefont {C.-A.}\ \bibnamefont
  {Pillet}},\ }\href {\doibase 10.1007/s00220-017-2947-1} {\bibfield  {journal}
  {\bibinfo  {journal} {Communications in Mathematical Physics}\ }\textbf
  {\bibinfo {volume} {357}},\ \bibinfo {pages} {77} (\bibinfo {year}
  {2017})}\BibitemShut {NoStop}%
\bibitem [{\citenamefont {Elouard}\ \emph
  {et~al.}(2017{\natexlab{b}})\citenamefont {Elouard}, \citenamefont
  {Herrera-Mart\'{\i}}, \citenamefont {Huard},\ and\ \citenamefont
  {Auff\`eves}}]{elou17_extracting}%
  \BibitemOpen
  \bibfield  {author} {\bibinfo {author} {\bibfnamefont {C.}~\bibnamefont
  {Elouard}}, \bibinfo {author} {\bibfnamefont {D.}~\bibnamefont
  {Herrera-Mart\'{\i}}}, \bibinfo {author} {\bibfnamefont {B.}~\bibnamefont
  {Huard}}, \ and\ \bibinfo {author} {\bibfnamefont {A.}~\bibnamefont
  {Auff\`eves}},\ }\href {\doibase 10.1103/PhysRevLett.118.260603} {\bibfield
  {journal} {\bibinfo  {journal} {Phys. Rev. Lett.}\ }\textbf {\bibinfo
  {volume} {118}},\ \bibinfo {pages} {260603} (\bibinfo {year}
  {2017}{\natexlab{b}})}\BibitemShut {NoStop}%
\bibitem [{\citenamefont {Elouard}\ and\ \citenamefont
  {Jordan}(2018)}]{elou18}%
  \BibitemOpen
  \bibfield  {author} {\bibinfo {author} {\bibfnamefont {C.}~\bibnamefont
  {Elouard}}\ and\ \bibinfo {author} {\bibfnamefont {A.~N.}\ \bibnamefont
  {Jordan}},\ }\href {\doibase 10.1103/PhysRevLett.120.260601} {\bibfield
  {journal} {\bibinfo  {journal} {Phys. Rev. Lett.}\ }\textbf {\bibinfo
  {volume} {120}},\ \bibinfo {pages} {260601} (\bibinfo {year}
  {2018})}\BibitemShut {NoStop}%
\bibitem [{\citenamefont {Buffoni}\ \emph {et~al.}(2019)\citenamefont
  {Buffoni}, \citenamefont {Solfanelli}, \citenamefont {Verrucchi},
  \citenamefont {Cuccoli},\ and\ \citenamefont {Campisi}}]{buff18}%
  \BibitemOpen
  \bibfield  {author} {\bibinfo {author} {\bibfnamefont {L.}~\bibnamefont
  {Buffoni}}, \bibinfo {author} {\bibfnamefont {A.}~\bibnamefont {Solfanelli}},
  \bibinfo {author} {\bibfnamefont {P.}~\bibnamefont {Verrucchi}}, \bibinfo
  {author} {\bibfnamefont {A.}~\bibnamefont {Cuccoli}}, \ and\ \bibinfo
  {author} {\bibfnamefont {M.}~\bibnamefont {Campisi}},\ }\href {\doibase
  10.1103/PhysRevLett.122.070603} {\bibfield  {journal} {\bibinfo  {journal}
  {Phys. Rev. Lett.}\ }\textbf {\bibinfo {volume} {122}},\ \bibinfo {pages}
  {070603} (\bibinfo {year} {2019})}\BibitemShut {NoStop}%
\bibitem [{\citenamefont {Esposito}\ \emph {et~al.}(2009)\citenamefont
  {Esposito}, \citenamefont {Harbola},\ and\ \citenamefont {Mukamel}}]{espo09}%
  \BibitemOpen
  \bibfield  {author} {\bibinfo {author} {\bibfnamefont {M.}~\bibnamefont
  {Esposito}}, \bibinfo {author} {\bibfnamefont {U.}~\bibnamefont {Harbola}}, \
  and\ \bibinfo {author} {\bibfnamefont {S.}~\bibnamefont {Mukamel}},\ }\href
  {\doibase 10.1103/RevModPhys.81.1665} {\bibfield  {journal} {\bibinfo
  {journal} {Rev. Mod. Phys.}\ }\textbf {\bibinfo {volume} {81}},\ \bibinfo
  {pages} {1665} (\bibinfo {year} {2009})}\BibitemShut {NoStop}%
\bibitem [{\citenamefont {Elouard}\ \emph
  {et~al.}(2017{\natexlab{c}})\citenamefont {Elouard}, \citenamefont
  {Herrera-Mart\'i}, \citenamefont {Clusel},\ and\ \citenamefont
  {Auff\`eves}}]{elo17}%
  \BibitemOpen
  \bibfield  {author} {\bibinfo {author} {\bibfnamefont {C.}~\bibnamefont
  {Elouard}}, \bibinfo {author} {\bibfnamefont {D.~A.}\ \bibnamefont
  {Herrera-Mart\'i}}, \bibinfo {author} {\bibfnamefont {M.}~\bibnamefont
  {Clusel}}, \ and\ \bibinfo {author} {\bibfnamefont {A.}~\bibnamefont
  {Auff\`eves}},\ }\href {\doibase 10.1038/s41534-017-0008-4} {\bibfield
  {journal} {\bibinfo  {journal} {npj Quantum Information}\ }\textbf {\bibinfo
  {volume} {3}},\ \bibinfo {pages} {9} (\bibinfo {year}
  {2017}{\natexlab{c}})}\BibitemShut {NoStop}%
\bibitem [{\citenamefont {{Naghiloo}}\ \emph
  {et~al.}(2017{\natexlab{b}})\citenamefont {{Naghiloo}}, \citenamefont
  {{Tan}}, \citenamefont {{Harrington}}, \citenamefont {{Alonso}},
  \citenamefont {{Lutz}}, \citenamefont {{Romito}},\ and\ \citenamefont
  {{Murch}}}]{nagh17}%
  \BibitemOpen
  \bibfield  {author} {\bibinfo {author} {\bibfnamefont {M.}~\bibnamefont
  {{Naghiloo}}}, \bibinfo {author} {\bibfnamefont {D.}~\bibnamefont {{Tan}}},
  \bibinfo {author} {\bibfnamefont {P.~M.}\ \bibnamefont {{Harrington}}},
  \bibinfo {author} {\bibfnamefont {J.~J.}\ \bibnamefont {{Alonso}}}, \bibinfo
  {author} {\bibfnamefont {E.}~\bibnamefont {{Lutz}}}, \bibinfo {author}
  {\bibfnamefont {A.}~\bibnamefont {{Romito}}}, \ and\ \bibinfo {author}
  {\bibfnamefont {K.~W.}\ \bibnamefont {{Murch}}},\ }\href@noop {} {\bibfield
  {journal} {\bibinfo  {journal} {arXiv:1703.05885}\ } (\bibinfo {year}
  {2017}{\natexlab{b}})}\BibitemShut {NoStop}%
\bibitem [{\citenamefont {Schulman}(1997)}]{Bookschulman}%
  \BibitemOpen
  \bibfield  {author} {\bibinfo {author} {\bibfnamefont {L.~S.}\ \bibnamefont
  {Schulman}},\ }\href@noop {} {\emph {\bibinfo {title} {{Time's Arrow and
  Quantum Measurement}}}}\ (\bibinfo  {publisher} {Cambridge University Press,
  Cambridge},\ \bibinfo {year} {1997})\BibitemShut {NoStop}%
\bibitem [{\citenamefont {Dressel}\ \emph {et~al.}(2017)\citenamefont
  {Dressel}, \citenamefont {Chantasri}, \citenamefont {Jordan},\ and\
  \citenamefont {Korotkov}}]{Dressel2017}%
  \BibitemOpen
  \bibfield  {author} {\bibinfo {author} {\bibfnamefont {J.}~\bibnamefont
  {Dressel}}, \bibinfo {author} {\bibfnamefont {A.}~\bibnamefont {Chantasri}},
  \bibinfo {author} {\bibfnamefont {A.~N.}\ \bibnamefont {Jordan}}, \ and\
  \bibinfo {author} {\bibfnamefont {A.~N.}\ \bibnamefont {Korotkov}},\ }\href
  {\doibase 10.1103/PhysRevLett.119.220507} {\bibfield  {journal} {\bibinfo
  {journal} {Phys. Rev. Lett.}\ }\textbf {\bibinfo {volume} {119}},\ \bibinfo
  {pages} {220507} (\bibinfo {year} {2017})}\BibitemShut {NoStop}%
\bibitem [{\citenamefont {Micadei}\ \emph {et~al.}(2017)\citenamefont
  {Micadei}, \citenamefont {Peterson}, \citenamefont {Souza}, \citenamefont
  {Sarthour}, \citenamefont {Oliveira}, \citenamefont {Landi}, \citenamefont
  {ao}, \citenamefont {Serra},\ and\ \citenamefont {Lutz}}]{kaon17}%
  \BibitemOpen
  \bibfield  {author} {\bibinfo {author} {\bibfnamefont {K.}~\bibnamefont
  {Micadei}}, \bibinfo {author} {\bibfnamefont {J.~P.~S.}\ \bibnamefont
  {Peterson}}, \bibinfo {author} {\bibfnamefont {A.~M.}\ \bibnamefont {Souza}},
  \bibinfo {author} {\bibfnamefont {R.~S.}\ \bibnamefont {Sarthour}}, \bibinfo
  {author} {\bibfnamefont {I.~S.}\ \bibnamefont {Oliveira}}, \bibinfo {author}
  {\bibfnamefont {G.~T.}\ \bibnamefont {Landi}}, \bibinfo {author}
  {\bibfnamefont {T.~B.~B.}\ \bibnamefont {ao}}, \bibinfo {author}
  {\bibfnamefont {R.~M.}\ \bibnamefont {Serra}}, \ and\ \bibinfo {author}
  {\bibfnamefont {E.}~\bibnamefont {Lutz}},\ }\href@noop {} {\bibfield
  {journal} {\bibinfo  {journal} {arXiv:1711.03323}\ } (\bibinfo {year}
  {2017})}\BibitemShut {NoStop}%
\bibitem [{\citenamefont {Manikandan}\ \emph
  {et~al.}(2019{\natexlab{a}})\citenamefont {Manikandan}, \citenamefont
  {Elouard},\ and\ \citenamefont {Jordan}}]{sree18}%
  \BibitemOpen
  \bibfield  {author} {\bibinfo {author} {\bibfnamefont {S.~K.}\ \bibnamefont
  {Manikandan}}, \bibinfo {author} {\bibfnamefont {C.}~\bibnamefont {Elouard}},
  \ and\ \bibinfo {author} {\bibfnamefont {A.~N.}\ \bibnamefont {Jordan}},\
  }\href {\doibase 10.1103/PhysRevA.99.022117} {\bibfield  {journal} {\bibinfo
  {journal} {Phys. Rev. A}\ }\textbf {\bibinfo {volume} {99}},\ \bibinfo
  {pages} {022117} (\bibinfo {year} {2019}{\natexlab{a}})}\BibitemShut
  {NoStop}%
\bibitem [{\citenamefont {Koch}\ \emph {et~al.}(2007)\citenamefont {Koch},
  \citenamefont {Yu}, \citenamefont {Gambetta}, \citenamefont {Houck},
  \citenamefont {Schuster}, \citenamefont {Majer}, \citenamefont {Blais},
  \citenamefont {Devoret}, \citenamefont {Girvin},\ and\ \citenamefont
  {Schoelkopf}}]{koch07}%
  \BibitemOpen
  \bibfield  {author} {\bibinfo {author} {\bibfnamefont {J.}~\bibnamefont
  {Koch}}, \bibinfo {author} {\bibfnamefont {T.~M.}\ \bibnamefont {Yu}},
  \bibinfo {author} {\bibfnamefont {J.}~\bibnamefont {Gambetta}}, \bibinfo
  {author} {\bibfnamefont {A.~A.}\ \bibnamefont {Houck}}, \bibinfo {author}
  {\bibfnamefont {D.~I.}\ \bibnamefont {Schuster}}, \bibinfo {author}
  {\bibfnamefont {J.}~\bibnamefont {Majer}}, \bibinfo {author} {\bibfnamefont
  {A.}~\bibnamefont {Blais}}, \bibinfo {author} {\bibfnamefont {M.~H.}\
  \bibnamefont {Devoret}}, \bibinfo {author} {\bibfnamefont {S.~M.}\
  \bibnamefont {Girvin}}, \ and\ \bibinfo {author} {\bibfnamefont {R.~J.}\
  \bibnamefont {Schoelkopf}},\ }\href {\doibase 10.1103/PhysRevA.76.042319}
  {\bibfield  {journal} {\bibinfo  {journal} {Phys. Rev. A}\ }\textbf {\bibinfo
  {volume} {76}},\ \bibinfo {pages} {042319} (\bibinfo {year}
  {2007})}\BibitemShut {NoStop}%
\bibitem [{\citenamefont {Paik}\ \emph {et~al.}(2011)\citenamefont {Paik},
  \citenamefont {Schuster}, \citenamefont {Bishop}, \citenamefont {Kirchmair},
  \citenamefont {Catelani}, \citenamefont {Sears}, \citenamefont {Johnson},
  \citenamefont {Reagor}, \citenamefont {Frunzio}, \citenamefont {Glazman},
  \citenamefont {Girvin}, \citenamefont {Devoret},\ and\ \citenamefont
  {Schoelkopf}}]{paik113D}%
  \BibitemOpen
  \bibfield  {author} {\bibinfo {author} {\bibfnamefont {H.}~\bibnamefont
  {Paik}}, \bibinfo {author} {\bibfnamefont {D.~I.}\ \bibnamefont {Schuster}},
  \bibinfo {author} {\bibfnamefont {L.~S.}\ \bibnamefont {Bishop}}, \bibinfo
  {author} {\bibfnamefont {G.}~\bibnamefont {Kirchmair}}, \bibinfo {author}
  {\bibfnamefont {G.}~\bibnamefont {Catelani}}, \bibinfo {author}
  {\bibfnamefont {A.~P.}\ \bibnamefont {Sears}}, \bibinfo {author}
  {\bibfnamefont {B.~R.}\ \bibnamefont {Johnson}}, \bibinfo {author}
  {\bibfnamefont {M.~J.}\ \bibnamefont {Reagor}}, \bibinfo {author}
  {\bibfnamefont {L.}~\bibnamefont {Frunzio}}, \bibinfo {author} {\bibfnamefont
  {L.~I.}\ \bibnamefont {Glazman}}, \bibinfo {author} {\bibfnamefont {S.~M.}\
  \bibnamefont {Girvin}}, \bibinfo {author} {\bibfnamefont {M.~H.}\
  \bibnamefont {Devoret}}, \ and\ \bibinfo {author} {\bibfnamefont {R.~J.}\
  \bibnamefont {Schoelkopf}},\ }\href@noop {} {\bibfield  {journal} {\bibinfo
  {journal} {Phys. Rev. Lett.}\ }\textbf {\bibinfo {volume} {107}},\ \bibinfo
  {pages} {240501} (\bibinfo {year} {2011})}\BibitemShut {NoStop}%
\bibitem [{\citenamefont {Castellanos-Beltran}\ \emph
  {et~al.}(2008)\citenamefont {Castellanos-Beltran}, \citenamefont {Irwin},
  \citenamefont {Hilton}, \citenamefont {Vale},\ and\ \citenamefont
  {Lehnert}}]{cast08}%
  \BibitemOpen
  \bibfield  {author} {\bibinfo {author} {\bibfnamefont {M.~A.}\ \bibnamefont
  {Castellanos-Beltran}}, \bibinfo {author} {\bibfnamefont {K.~D.}\
  \bibnamefont {Irwin}}, \bibinfo {author} {\bibfnamefont {G.~C.}\ \bibnamefont
  {Hilton}}, \bibinfo {author} {\bibfnamefont {L.~R.}\ \bibnamefont {Vale}}, \
  and\ \bibinfo {author} {\bibfnamefont {K.~W.}\ \bibnamefont {Lehnert}},\
  }\href@noop {} {\bibfield  {journal} {\bibinfo  {journal} {Nature Physics}\
  }\textbf {\bibinfo {volume} {4}},\ \bibinfo {pages} {929} (\bibinfo {year}
  {2008})}\BibitemShut {NoStop}%
\bibitem [{\citenamefont {Hatridge}\ \emph {et~al.}(2011)\citenamefont
  {Hatridge}, \citenamefont {Vijay}, \citenamefont {Slichter}, \citenamefont
  {Clarke},\ and\ \citenamefont {Siddiqi}}]{hatr11para}%
  \BibitemOpen
  \bibfield  {author} {\bibinfo {author} {\bibfnamefont {M.}~\bibnamefont
  {Hatridge}}, \bibinfo {author} {\bibfnamefont {R.}~\bibnamefont {Vijay}},
  \bibinfo {author} {\bibfnamefont {D.~H.}\ \bibnamefont {Slichter}}, \bibinfo
  {author} {\bibfnamefont {J.}~\bibnamefont {Clarke}}, \ and\ \bibinfo {author}
  {\bibfnamefont {I.}~\bibnamefont {Siddiqi}},\ }\href {\doibase
  10.1103/PhysRevB.83.134501} {\bibfield  {journal} {\bibinfo  {journal} {Phys.
  Rev. B}\ }\textbf {\bibinfo {volume} {83}},\ \bibinfo {pages} {134501}
  (\bibinfo {year} {2011})}\BibitemShut {NoStop}%
\bibitem [{\citenamefont {Breuer}\ and\ \citenamefont
  {Petruccione}(2007)}]{breu07}%
  \BibitemOpen
  \bibfield  {author} {\bibinfo {author} {\bibfnamefont {H.}~\bibnamefont
  {Breuer}}\ and\ \bibinfo {author} {\bibfnamefont {F.}~\bibnamefont
  {Petruccione}},\ }\href@noop {} {\emph {\bibinfo {title} {The Theory of Open
  Quantum Systems}}}\ (\bibinfo  {publisher} {OUP Oxford},\ \bibinfo {year}
  {2007})\BibitemShut {NoStop}%
\bibitem [{\citenamefont {Jacobs}\ and\ \citenamefont {Steck}(2006)}]{jaco06}%
  \BibitemOpen
  \bibfield  {author} {\bibinfo {author} {\bibfnamefont {K.}~\bibnamefont
  {Jacobs}}\ and\ \bibinfo {author} {\bibfnamefont {D.~A.}\ \bibnamefont
  {Steck}},\ }\href@noop {} {\bibfield  {journal} {\bibinfo  {journal}
  {Contemp. Phys.}\ }\textbf {\bibinfo {volume} {47}},\ \bibinfo {pages} {279}
  (\bibinfo {year} {2006})}\BibitemShut {NoStop}%
\bibitem [{\citenamefont {Korotkov}\ and\ \citenamefont
  {Jordan}(2006)}]{koro06}%
  \BibitemOpen
  \bibfield  {author} {\bibinfo {author} {\bibfnamefont {A.~N.}\ \bibnamefont
  {Korotkov}}\ and\ \bibinfo {author} {\bibfnamefont {A.~N.}\ \bibnamefont
  {Jordan}},\ }\href {\doibase 10.1103/PhysRevLett.97.166805} {\bibfield
  {journal} {\bibinfo  {journal} {Phys. Rev. Lett.}\ }\textbf {\bibinfo
  {volume} {97}},\ \bibinfo {pages} {166805} (\bibinfo {year}
  {2006})}\BibitemShut {NoStop}%
\bibitem [{\citenamefont {Katz}\ \emph {et~al.}(2006)\citenamefont {Katz},
  \citenamefont {Ansmann}, \citenamefont {Bialczak}, \citenamefont {Lucero},
  \citenamefont {McDermott}, \citenamefont {Neeley}, \citenamefont {Steffen},
  \citenamefont {Weig}, \citenamefont {Cleland}, \citenamefont {Martinis},\
  and\ \citenamefont {Korotkov}}]{katz06}%
  \BibitemOpen
  \bibfield  {author} {\bibinfo {author} {\bibfnamefont {N.}~\bibnamefont
  {Katz}}, \bibinfo {author} {\bibfnamefont {M.}~\bibnamefont {Ansmann}},
  \bibinfo {author} {\bibfnamefont {R.~C.}\ \bibnamefont {Bialczak}}, \bibinfo
  {author} {\bibfnamefont {E.}~\bibnamefont {Lucero}}, \bibinfo {author}
  {\bibfnamefont {R.}~\bibnamefont {McDermott}}, \bibinfo {author}
  {\bibfnamefont {M.}~\bibnamefont {Neeley}}, \bibinfo {author} {\bibfnamefont
  {M.}~\bibnamefont {Steffen}}, \bibinfo {author} {\bibfnamefont {E.~M.}\
  \bibnamefont {Weig}}, \bibinfo {author} {\bibfnamefont {A.~N.}\ \bibnamefont
  {Cleland}}, \bibinfo {author} {\bibfnamefont {J.~M.}\ \bibnamefont
  {Martinis}}, \ and\ \bibinfo {author} {\bibfnamefont {A.~N.}\ \bibnamefont
  {Korotkov}},\ }\href {\doibase 10.1126/science.1126475} {\bibfield  {journal}
  {\bibinfo  {journal} {Science}\ }\textbf {\bibinfo {volume} {312}},\ \bibinfo
  {pages} {1498} (\bibinfo {year} {2006})}\BibitemShut {NoStop}%
\bibitem [{\citenamefont {Katz}\ \emph {et~al.}(2008)\citenamefont {Katz},
  \citenamefont {Neeley}, \citenamefont {Ansmann}, \citenamefont {Bialczak},
  \citenamefont {Hofheinz}, \citenamefont {Lucero}, \citenamefont {O'Connell},
  \citenamefont {Wang}, \citenamefont {Cleland}, \citenamefont {Martinis},\
  and\ \citenamefont {Korotkov}}]{katz08}%
  \BibitemOpen
  \bibfield  {author} {\bibinfo {author} {\bibfnamefont {N.}~\bibnamefont
  {Katz}}, \bibinfo {author} {\bibfnamefont {M.}~\bibnamefont {Neeley}},
  \bibinfo {author} {\bibfnamefont {M.}~\bibnamefont {Ansmann}}, \bibinfo
  {author} {\bibfnamefont {R.~C.}\ \bibnamefont {Bialczak}}, \bibinfo {author}
  {\bibfnamefont {M.}~\bibnamefont {Hofheinz}}, \bibinfo {author}
  {\bibfnamefont {E.}~\bibnamefont {Lucero}}, \bibinfo {author} {\bibfnamefont
  {A.}~\bibnamefont {O'Connell}}, \bibinfo {author} {\bibfnamefont
  {H.}~\bibnamefont {Wang}}, \bibinfo {author} {\bibfnamefont {A.~N.}\
  \bibnamefont {Cleland}}, \bibinfo {author} {\bibfnamefont {J.~M.}\
  \bibnamefont {Martinis}}, \ and\ \bibinfo {author} {\bibfnamefont {A.~N.}\
  \bibnamefont {Korotkov}},\ }\href {\doibase 10.1103/PhysRevLett.101.200401}
  {\bibfield  {journal} {\bibinfo  {journal} {Phys. Rev. Lett.}\ }\textbf
  {\bibinfo {volume} {101}},\ \bibinfo {pages} {200401} (\bibinfo {year}
  {2008})}\BibitemShut {NoStop}%
\bibitem [{\citenamefont {Kim}\ \emph {et~al.}(2011)\citenamefont {Kim},
  \citenamefont {Lee}, \citenamefont {Kwon},\ and\ \citenamefont
  {Kim}}]{kim11}%
  \BibitemOpen
  \bibfield  {author} {\bibinfo {author} {\bibfnamefont {Y.-S.}\ \bibnamefont
  {Kim}}, \bibinfo {author} {\bibfnamefont {J.-C.}\ \bibnamefont {Lee}},
  \bibinfo {author} {\bibfnamefont {O.}~\bibnamefont {Kwon}}, \ and\ \bibinfo
  {author} {\bibfnamefont {Y.-H.}\ \bibnamefont {Kim}},\ }\href@noop {}
  {\bibfield  {journal} {\bibinfo  {journal} {Nature Physics}\ }\textbf
  {\bibinfo {volume} {8}},\ \bibinfo {pages} {117} (\bibinfo {year}
  {2011})}\BibitemShut {NoStop}%
\bibitem [{\citenamefont {Paraoanu}(2011)}]{para11}%
  \BibitemOpen
  \bibfield  {author} {\bibinfo {author} {\bibfnamefont {G.~S.}\ \bibnamefont
  {Paraoanu}},\ }\href {\doibase 10.1007/s10701-011-9542-7} {\bibfield
  {journal} {\bibinfo  {journal} {Foundations of Physics}\ }\textbf {\bibinfo
  {volume} {41}},\ \bibinfo {pages} {1214} (\bibinfo {year}
  {2011})}\BibitemShut {NoStop}%
\bibitem [{SM()}]{SM}%
  \BibitemOpen
  \href@noop {} {}\bibinfo {note} {See Supplemental Material}\BibitemShut
  {NoStop}%
\bibitem [{not()}]{note}%
  \BibitemOpen
  \href@noop {} {}\bibinfo {note} {The quantum efficiency includes the combined
  effects of collection efficiency, added noise from the amplification chain
  and nearly negligible environmental dephasing characterized by $T_2^* = 15\
  \mu$s.}\BibitemShut {Stop}%
\bibitem [{\citenamefont {Korotkov}(2011)}]{koro11}%
  \BibitemOpen
  \bibfield  {author} {\bibinfo {author} {\bibfnamefont {A.~N.}\ \bibnamefont
  {Korotkov}},\ }\href@noop {} {\bibfield  {journal} {\bibinfo  {journal}
  {arXiv:1111.4016}\ } (\bibinfo {year} {2011})}\BibitemShut {NoStop}%
\bibitem [{\citenamefont {Hatridge}\ \emph {et~al.}(2013)\citenamefont
  {Hatridge}, \citenamefont {Shankar}, \citenamefont {Mirrahimi}, \citenamefont
  {Schackert}, \citenamefont {Geerlings}, \citenamefont {Brecht}, \citenamefont
  {Sliwa}, \citenamefont {Abdo}, \citenamefont {Frunzio}, \citenamefont
  {Girvin}, \citenamefont {Schoelkopf},\ and\ \citenamefont
  {Devoret}}]{hatr13}%
  \BibitemOpen
  \bibfield  {author} {\bibinfo {author} {\bibfnamefont {M.}~\bibnamefont
  {Hatridge}}, \bibinfo {author} {\bibfnamefont {S.}~\bibnamefont {Shankar}},
  \bibinfo {author} {\bibfnamefont {M.}~\bibnamefont {Mirrahimi}}, \bibinfo
  {author} {\bibfnamefont {F.}~\bibnamefont {Schackert}}, \bibinfo {author}
  {\bibfnamefont {K.}~\bibnamefont {Geerlings}}, \bibinfo {author}
  {\bibfnamefont {T.}~\bibnamefont {Brecht}}, \bibinfo {author} {\bibfnamefont
  {K.~M.}\ \bibnamefont {Sliwa}}, \bibinfo {author} {\bibfnamefont
  {B.}~\bibnamefont {Abdo}}, \bibinfo {author} {\bibfnamefont {L.}~\bibnamefont
  {Frunzio}}, \bibinfo {author} {\bibfnamefont {S.~M.}\ \bibnamefont {Girvin}},
  \bibinfo {author} {\bibfnamefont {R.~J.}\ \bibnamefont {Schoelkopf}}, \ and\
  \bibinfo {author} {\bibfnamefont {M.~H.}\ \bibnamefont {Devoret}},\
  }\href@noop {} {\bibfield  {journal} {\bibinfo  {journal} {Science}\ }\textbf
  {\bibinfo {volume} {339}},\ \bibinfo {pages} {178} (\bibinfo {year}
  {2013})}\BibitemShut {NoStop}%
\bibitem [{\citenamefont {Jarzynski}(1999)}]{jarz99}%
  \BibitemOpen
  \bibfield  {author} {\bibinfo {author} {\bibfnamefont {C.}~\bibnamefont
  {Jarzynski}},\ }\href@noop {} {\bibfield  {journal} {\bibinfo  {journal}
  {Journal of Statistical Physics}\ }\textbf {\bibinfo {volume} {96}},\
  \bibinfo {pages} {415} (\bibinfo {year} {1999})}\BibitemShut {NoStop}%
\bibitem [{\citenamefont {Murashita}\ \emph {et~al.}(2014)\citenamefont
  {Murashita}, \citenamefont {Funo},\ and\ \citenamefont {Ueda}}]{mura14}%
  \BibitemOpen
  \bibfield  {author} {\bibinfo {author} {\bibfnamefont {Y.}~\bibnamefont
  {Murashita}}, \bibinfo {author} {\bibfnamefont {K.}~\bibnamefont {Funo}}, \
  and\ \bibinfo {author} {\bibfnamefont {M.}~\bibnamefont {Ueda}},\ }\href
  {\doibase 10.1103/PhysRevE.90.042110} {\bibfield  {journal} {\bibinfo
  {journal} {Phys. Rev. E}\ }\textbf {\bibinfo {volume} {90}},\ \bibinfo
  {pages} {042110} (\bibinfo {year} {2014})}\BibitemShut {NoStop}%
\bibitem [{\citenamefont {Funo}\ \emph {et~al.}(2015)\citenamefont {Funo},
  \citenamefont {Murashita},\ and\ \citenamefont {Ueda}}]{funo15}%
  \BibitemOpen
  \bibfield  {author} {\bibinfo {author} {\bibfnamefont {K.}~\bibnamefont
  {Funo}}, \bibinfo {author} {\bibfnamefont {Y.}~\bibnamefont {Murashita}}, \
  and\ \bibinfo {author} {\bibfnamefont {M.}~\bibnamefont {Ueda}},\ }\href
  {\doibase 10.1088/1367-2630/17/7/075005} {\bibfield  {journal} {\bibinfo
  {journal} {New Journal of Physics}\ }\textbf {\bibinfo {volume} {17}},\
  \bibinfo {pages} {075005} (\bibinfo {year} {2015})}\BibitemShut {NoStop}%
\bibitem [{\citenamefont {Manzano}\ \emph {et~al.}(2015)\citenamefont
  {Manzano}, \citenamefont {Horowitz},\ and\ \citenamefont
  {Parrondo}}]{manz15}%
  \BibitemOpen
  \bibfield  {author} {\bibinfo {author} {\bibfnamefont {G.}~\bibnamefont
  {Manzano}}, \bibinfo {author} {\bibfnamefont {J.~M.}\ \bibnamefont
  {Horowitz}}, \ and\ \bibinfo {author} {\bibfnamefont {J.~M.~R.}\ \bibnamefont
  {Parrondo}},\ }\href {\doibase 10.1103/PhysRevE.92.032129} {\bibfield
  {journal} {\bibinfo  {journal} {Phys. Rev. E}\ }\textbf {\bibinfo {volume}
  {92}},\ \bibinfo {pages} {032129} (\bibinfo {year} {2015})}\BibitemShut
  {NoStop}%
\bibitem [{\citenamefont {Manikandan}\ \emph
  {et~al.}(2019{\natexlab{b}})\citenamefont {Manikandan}, \citenamefont
  {Elouard},\ and\ \citenamefont {Jordan}}]{sree19}%
  \BibitemOpen
  \bibfield  {author} {\bibinfo {author} {\bibfnamefont {S.~K.}\ \bibnamefont
  {Manikandan}}, \bibinfo {author} {\bibfnamefont {C.}~\bibnamefont {Elouard}},
  \ and\ \bibinfo {author} {\bibfnamefont {A.~N.}\ \bibnamefont {Jordan}},\
  }\href {\doibase 10.1103/PhysRevA.99.022117} {\bibfield  {journal} {\bibinfo
  {journal} {Phys. Rev. A}\ }\textbf {\bibinfo {volume} {99}},\ \bibinfo
  {pages} {022117} (\bibinfo {year} {2019}{\natexlab{b}})}\BibitemShut
  {NoStop}%
  \bibitem[{\citenamefont{{Jordan} and {Korotkov}}(2006)}]{jord06}
\bibinfo{author}{\bibfnamefont{A.~N.} \bibnamefont{{Jordan}}} \bibnamefont{and}
  \bibinfo{author}{\bibfnamefont{A.~N.} \bibnamefont{{Korotkov}}},
  \bibinfo{journal}{Physical Review B} \textbf{\bibinfo{volume}{74}},
  \bibinfo{eid}{085307} (\bibinfo{year}{2006}).
\end{thebibliography}

\end{document}